 \documentclass{vgtc}                          

\graphicspath{{figs/}{pictures/}{images/}{./}} 

\usepackage{times}                     

\usepackage{tabu}                      
\usepackage{booktabs}                  
\usepackage{lipsum}                    
\usepackage{mwe}                       

\usepackage{mathptmx}                  

\usepackage{tabularray}
\usepackage{wrapfig}
\usepackage[svgnames]{xcolor}
\usepackage[basic]{complexity}
\usepackage{amsmath}
\usepackage{amssymb}
\usepackage{mathtools}
\usepackage{csquotes}
\usepackage{float}

\usepackage{subcaption}
\usepackage[pagebackref,bookmarks]{hyperref}
\usepackage[nameinlink,capitalise]{cleveref}

\onlineid{1382}

\vgtccategory{Research}

\vgtcinsertpkg

\title{Adaptive Uncertainty Visualization With jk-Plots}

\author{Nikolaus Piccolotto\thanks{e-mail: \{firstname.lastname\}@\{tuwien.ac.at, univie.ac.at\}}\\ %
        \scriptsize TU Wien %
\and Daniel Pahr\\ %
     \scriptsize University of Vienna %
 \and Laura Lotteraner\\ %
     \scriptsize University of Vienna %
\and Fatih Öztank\\ %
     \scriptsize TU Wien %
\and Markus Bögl\\ %
     \scriptsize TU Wien}

\abstract{

An important task of uncertainty visualizations is to communicate how the number of available observations influences the precision of estimates, such as a Normal distribution's mean. 
In this paper, we propose an adaptive approach to visualizing univariate data that produces salient representations while highlighting the uncertainty introduced by the sample size. 
We introduce jk-plots, inspired by educational material of Bayesian statistics, that combine visualizations of plausible reference models (emulating popular uncertainty visualization idioms) and the observations. 
We describe a visualization pipeline to create jk-plots, which we implemented in an R package. 
We demonstrate jk-plots using synthetic datasets of varying size and discuss possible usage scenarios in domain expert interviews.
}


\begin{document}



\maketitle

\section{Introduction}

In statistical inference, the number of available observations crucially affects the uncertainty of estimates and the power of tests.
Viewing visualizations has been characterized as \enquote{graphical inference}~\cite{hullman2021,wickham2010}: The viewer implicitly assumes a reference model (e.g., a Normal distribution) and compares it to a visual representation of the dataset (e.g., a histogram), judging whether the data are consistent with the reference model.
Two problems associated with commonly employed uncertainty visualizations may hinder this task.

\begin{wrapfigure}{o}{0.2\textwidth}
\includegraphics[width=0.9\linewidth]{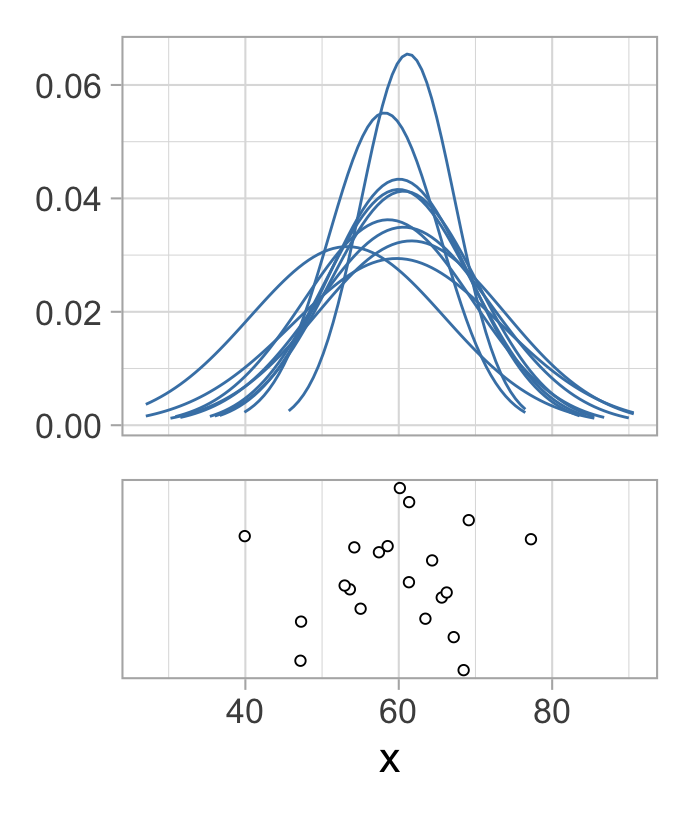} 
\caption{Kruschke-style.}
\label{fig:kruschke}
\end{wrapfigure}

First, the reference model may be ambiguous, as a sample may stem from a wide range of distributions (\cref{fig:kruschke}, top). The fewer observations, the more uncertain are the parameters of the reference model. This kind of uncertainty is often visually conveyed via certainty bounds of a parameter, e.g, confidence intervals or the standard error of the mean. However, certainty bounds hide the possible reference models and are thus reductive, may be misinterpreted~\cite{belia2005, correll2014}, may need special corrections for small samples, or may not exist for every parameter. 
On the other hand, showing the sample as-is, e.g., (\cref{fig:kruschke}, bottom), hides the certainty bounds of a parameter in addition to the reference models, and does not scale to large samples due to overplotting and occlusion. 
Therefore, popular uncertainty visualizations do not appropriately communicate the sample size's implications (\textbf{P1}).

The second problem arises when visualizations of empirical distributions appear broken with too few observations (\textbf{P2}). 
An empirical cumulative distribution function (eCDF), for example, can appear jagged and staircase-like (e.g., \cref{fig:comp:cont10:regular}), a histogram may reduce to a few disparate bars, and a boxplot to a single line. It stands to reason that readability of uncertainty visualizations would increase, if they seamlessly adapted their appearance. Semiotic uncertainty visualization approaches~\cite{correll2018,maceachren2012} would even encourage that unreliable estimates become unreadable.

We see potential in expanding the design space of uncertainty visualizations to address these problems by i) quantifying the uncertainty introduced by small samples, ii) encoding this uncertainty in an intuitive manner, and iii) seamlessly adapting the visual encoding to the uncertainty.


\setlength{\wrapoverhang}{4pt}%
\setlength{\columnsep}{2pt}%
\setlength{\intextsep}{4pt}%

We propose a technique based on plots by John K. Kruschke~\cite{kruschke2015} that show i) the observed data unsummarized, and ii) models that could realistically have generated the data, based on the Bayesian posterior. The reference of \texttt{ggdist}~\cite{kay2024} describes the creation of such plots, and refers to them as ``Kruschke-style'' visualizations (\cref{fig:kruschke}). In this paper, we investigate the feasibility of such plots by applying them to synthetic datasets and by interviewing domain experts.

Our contributions are i) the extension of Kruschke-style visualizations to other idioms and data, thus forming the \emph{jk-plot} family; ii) a proof-of-concept implementation in an R package, available in our \href{https://osf.io/wzv2g}{OSF repository}; iii) a visual comparison to empirical distribution visualizations; and iv) interviews with experts discussing how jk-plots would apply to real data.


\section{Related Work}
\label{sec:rw}


Visual representations of uncertain data, known as \emph{uncertainty visualizations}, are a central element of uncertainty and risk communication~\cite{spiegelhalter2017, vanderbles2019}, and have been studied for several years~\cite{bonneau2014, kamal2021, thomson2005}.

\paragraph*{Uncertainty Visualization.} Based on Padilla et al.~\cite{padilla2021} we distinguish four categories of uncertainty visualization: ensemble, summary, hybrid, and semiotic.
\textit{Ensembles} frame uncertainty as frequencies and display individual observations from a sample or draws from an analytical distribution. Examples include jittered dot plots (\cref{fig:kruschke}, bottom), blue noise plots~\cite{vanonzenoodt2021}, beeswarm plots, or relaxed dot plots~\cite{rodrigues2023}.
\textit{Summary visualizations}, on the other hand, present descriptive summary statistics from the samples or the analytical distribution, such as the quartiles encoded in the box of a boxplot~\cite{spear1952, tukey1977}. 
Other examples include quantile dot plots~\cite{kay2016}, probability density functions (PDF), or cumulative distribution functions (CDF).
When ensembles and summary visualizations are combined, the result is a \textit{hybrid} uncertainty visualization. The goal of the combination is that each visualization may ``defend'' against the others' weaknesses~\cite{correll2023}. Examples are raincloud plots~\cite{allen2019} and v-plots~\cite{blumenschein2020a}, which were proposed specifically to compare distributions.
Finally, \textit{semiotic} uncertainty visualizations~\cite{maceachren2012}, such as  value-suppressing uncertainty palettes~\cite{correll2018}, display encoded values with lower readability the more uncertain they are. 
In practice, libraries such as ggdist~\cite{kay2024} may be used to create custom versions of many of the examples mentioned above.
However, none of these approaches specifically target the visualization of uncertainty arising from small sample sizes.


\paragraph*{Sample Size in Statistics.} \emph{Summary} uncertainty visualizations mainly employ \emph{descriptive} statistics, which are computed directly from observations, not a statistical model of the underlying population or process. When such a statistical model is used, we speak of \emph{inferential} statistics. 
An important step in statistical analysis is determining the sample size required to achieve a given level of precision. In frequentist statistics, this process is known as \emph{power analysis}~\cite{cohen1988}. 
Tests may assume a minimum sample size themselves, conventionally set at around 30 observations. Frequentist inferential statistics, such as the standard error and the confidence interval, are often included in uncertainty visualizations, although the latter are frequently misinterpreted~\cite{belia2005, correll2014}.
The analog process in Bayesian statistics, known as \emph{sample size planning}~\cite[Chp.~13]{kruschke2015}, evaluates a chosen sample size by repeatedly simulating data under assumed parameter values, fitting the model to each simulated dataset, and checking whether the resulting posterior satisfies a defined criterion, e.g., whether the credible interval excludes a threshold. The proportion of simulated datasets that satisfy this criterion serves as the analog to power in frequentist analysis.
\enquote{Kruschke-style visualizations} (e.g.,~\cite[Fig.~19.3 on p.~563]{kruschke2015}, see also \cref{fig:kruschke}) were suggested for \emph{model checking}, i.e., a way to determine how well a fitted Bayesian model represents the data~\cite{gelman2020}. They depict the raw observations combined with a small number of draws from the model parameters' posterior distribution, represented by Kruschke as PDFs.
This visualization approach is closely related to \emph{posterior predictive checking}~\cite{gabry2019, gelman2020, vandeschoot2021}, an integral part of Bayesian data analysis that investigates whether the observations could have been generated by the model's posterior. The R package \texttt{bayesplot}~\cite{bayesplot} provides common plots needed for Bayesian model checking. Though the aim is identical, the main difference to jk-plots is that these graphics generally do not attempt to replicate familiar visualization idioms. 

\section{jk-Plots}
\label{sec:jkplot}

jk-plots (\cref{fig:kruschke}) display plausible reference models (top) that could have generated the observations (bottom). Their conceptual visualization pipeline consists of two steps: 1) Identify the reference models, and 2) visualize them appropriately together with the observations.
For \emph{reference model identification} (step~1), we employ Bayesian statistical inference to estimate likely model parameters, because its posterior provides a distribution and not just, e.g., an interval. Theoretically, this step could employ other means as well, e.g., uniformly sampling within the standard errors of the mean and standard deviation.
Our \emph{visualization} approaches (step~2) depend on the type of variable (categorical or continuous) and how many reference models were identified. For categorical data, we use the widely known pie chart idiom. Continuous data follow commonly employed uncertainty visualization idioms: histograms, box plots, cumulative distribution functions (CDFs), and probability density functions (PDFs). We identify the Bayesian posterior via Hamiltonian-Monte-Carlo sampling and collect several thousand draws from it. As each individual draw represents parameters for one plausible reference model, visual summarization becomes necessary. We implemented three common strategies: \emph{subsetting} randomly selects a user-defined number of models (\cref{fig:comp:cat}), \emph{bands} aggregates models into 50-80-95\% intervals (\cref{fig:comp:cont}), and \emph{gradient} applies a visual fade-out from the main trend (\cref{fig:comp:skew}). 

By combining a visual representation of the sample with visualizations of plausibly parameterized models (cf.~\cref{fig:kruschke}), jk-plots show the impact of the sample size threefold. First, the visual representation of the sample (\cref{fig:kruschke}, bottom) makes their count directly visible. Second, the precision of the posterior distribution is visible as the variance in the reference model visualization (\cref{fig:kruschke}, top). Third, when a suitable visualization idiom of the latter is used, parameter estimations may be read off the graphic directly. E.g., certainty bounds for mean estimates or any other cumulative probability may be read off CDFs with a \emph{band} summary. We refer readers to \cref{sec:examples} for example visualizations. In the following, we provide a more technical description of our jk-plot implementation. Readers may obtain the exact code used to generate figures in this paper in our \href{https://osf.io/wzv2g}{OSF repository}. 

\paragraph{Implementation.} jk-plots are based on Bayesian inference of model parameters. We refer the reader to accessible introductions (e.g., \cite{kruschke2015, mcelreath2022}) for detailed explanations, but offer some intuition here.
Suppose we toss a coin ten times to test whether it is fair, i.e., both sides come up with equal probability. 
To judge the result, we assume a model (independent tosses: Binomial distribution), and focus on the parameter of interest ($p =p(head)$). We compute the probability of the observed outcome (i.e., the ten tosses) for varying values of $p$, which yields the parameter's \emph{posterior distribution}. Its modes mark the most likely values, given the observed data and model, and its mass demarcates the interval of plausible values. 
Inspecting the posterior now allows us to judge the coin's fairness.
A Bayesian workflow generally consists of the steps \emph{formulate model}, \emph{set prior distributions}, \emph{estimate posterior}~\cite{vandeschoot2021}. We briefly describe how we implemented them.
We support \textbf{models} for categorical and continuous data, which are defined by the data type: Continuous data are by default modeled as a Normal distribution, while categorical data is expected to come from a categorical distribution. The model always estimates only an intercept without predictors ($y \sim 1$).
We aim to set uninformative \textbf{priors} as we cannot assume prior knowledge to inform them. For continuous data we set priors on z-scores. The prior for the mean ($\mu$) is a Normal(0, 50) distribution and an Exponential(1/100) for the standard deviation ($\sigma$). While the $\mu$ prior is centered at the sample mean, it is allowed to deviate by 100 sample standard deviations (as 95\% of probability in Normal(0,50) is within (-100,100)). Similarly, the prior for $\sigma$ allows values up to 370 times the observed sample value. We consider those priors sufficiently uninformative. While it is considered good practice to define priors without \enquote{double-dipping}~\cite{mcelreath2022, vandeschoot2021} (using the observations for setting priors \emph{and} estimating model parameters), data-driven priors are sometimes argued for on the grounds that if no prior knowledge exists it at least saves compute cycles. For categorical data, we use a logit-link function and set the priors for the non-reference categories to \emph{Normal(0, 20)}. 
To estimate the \textbf{posterior}, we scale continuous data to 0 mean and 1 standard deviation. We use the R package \texttt{brms} as an interface to the probabilistic programming language Stan~\cite{stan}. 1000 warm-up iterations out of 5000 in total with 4 parallel chains lead to $4 \times 4000$ available posterior draws.

\section{Case Study}
\label{sec:examples}

To identify the advantages and shortcomings of jk-plots, we compare them to their empirical counterparts for synthetic data derived from well-defined distributions.

\begin{figure}[tbhp]
    \centering

    \begin{subfigure}{0.3\linewidth}
        \includegraphics[width=0.9\linewidth]{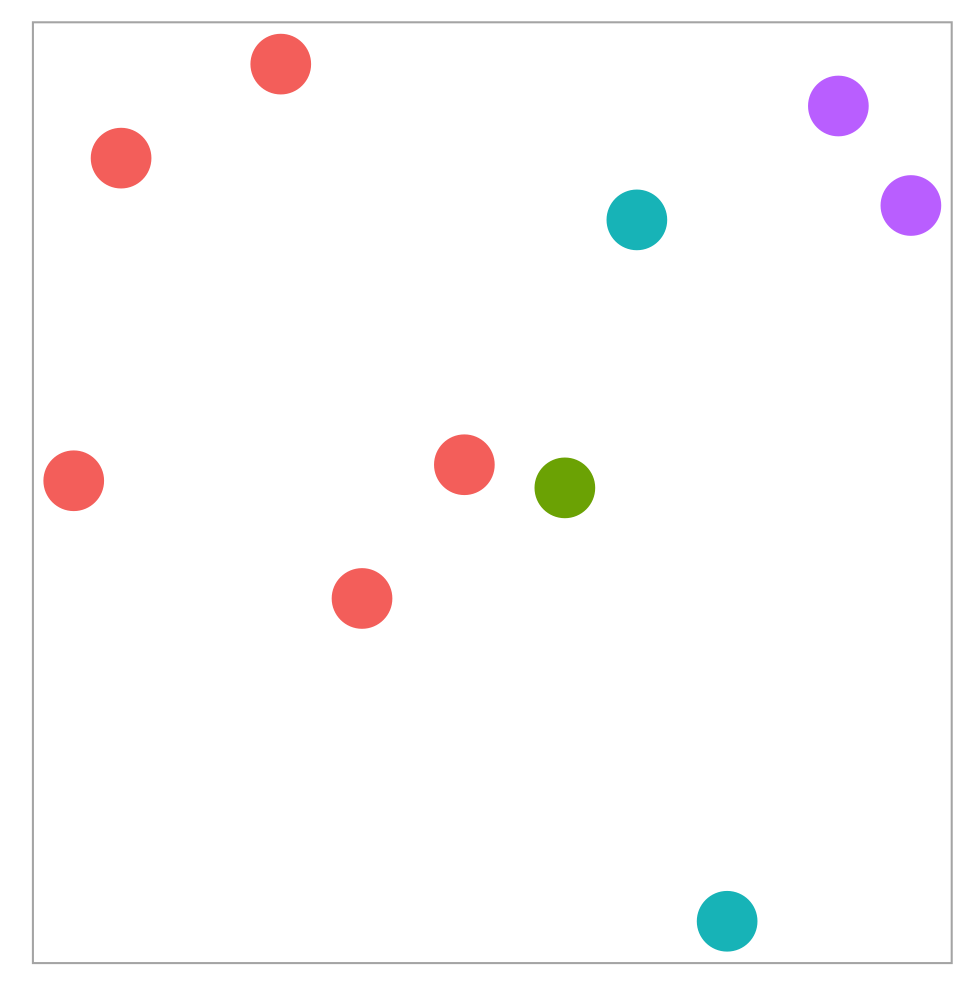} 
        \caption{10 Samples 
        }
        \label{fig:comp:cat10:sample}
    \end{subfigure}
    \begin{subfigure}{0.3\linewidth}
        \includegraphics[width=0.9\linewidth]{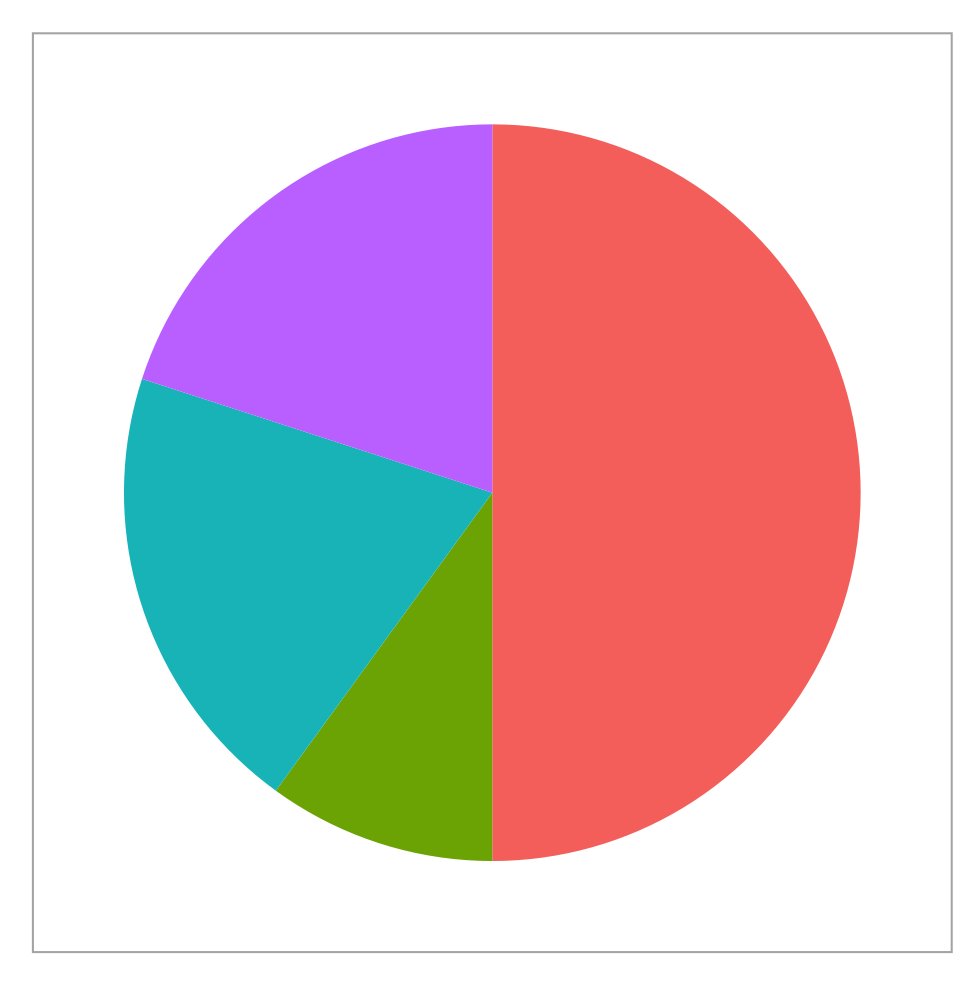} 
        \caption{Empirical Pie Chart}
        \label{fig:comp:cat10:regular}
    \end{subfigure}
    \begin{subfigure}{0.3\linewidth}
        \includegraphics[width=0.9\linewidth]{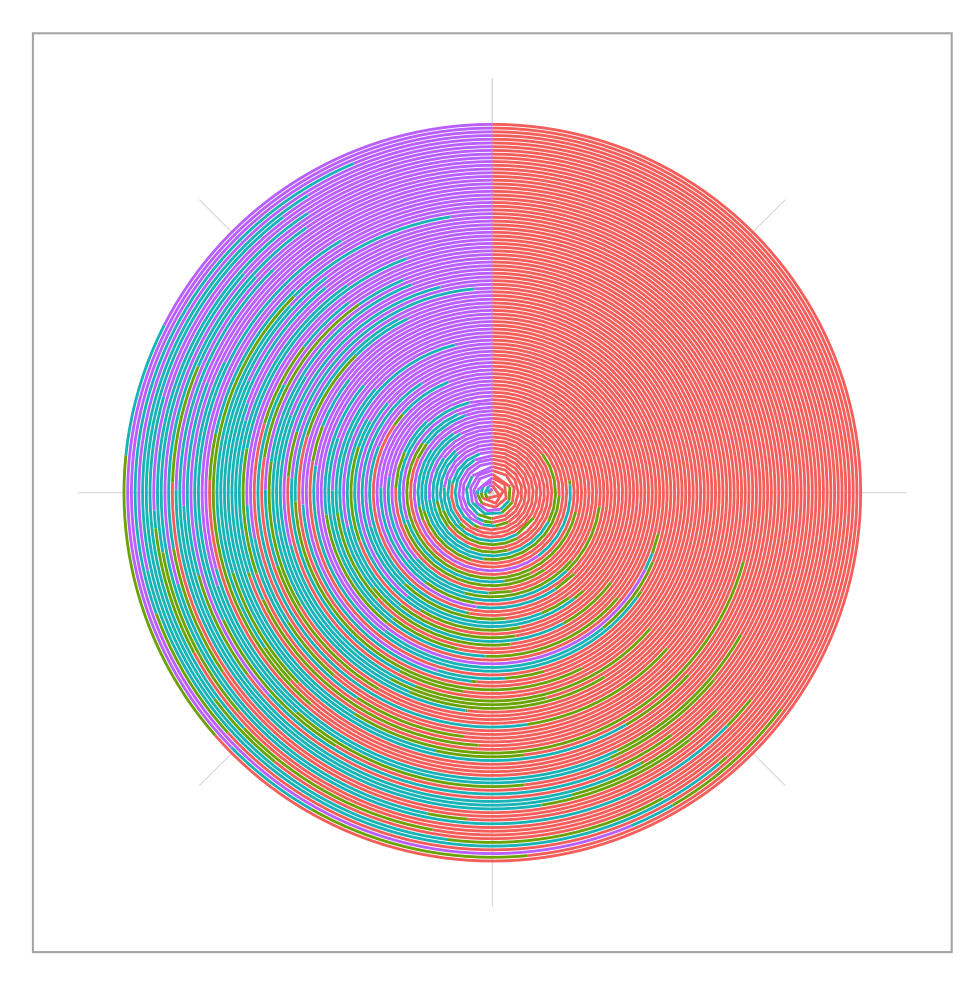} 
        \caption{jk Pie Chart}
        \label{fig:comp:cat10:ours}
    \end{subfigure}

    \begin{subfigure}{0.3\linewidth}
        \includegraphics[width=0.9\linewidth]{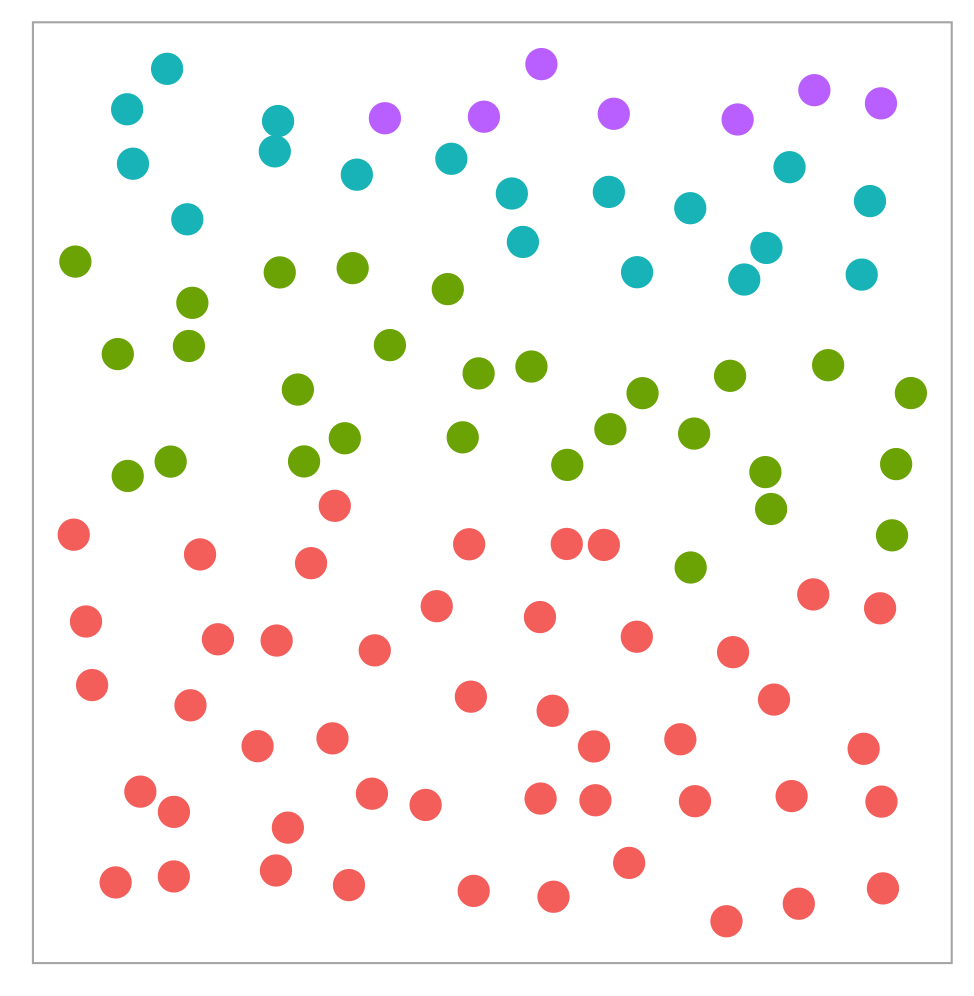} 
        \caption{100 Samples 
        }
        \label{fig:comp:cat100:sample}
    \end{subfigure}
    \begin{subfigure}{0.3\linewidth}
        \includegraphics[width=0.9\linewidth]{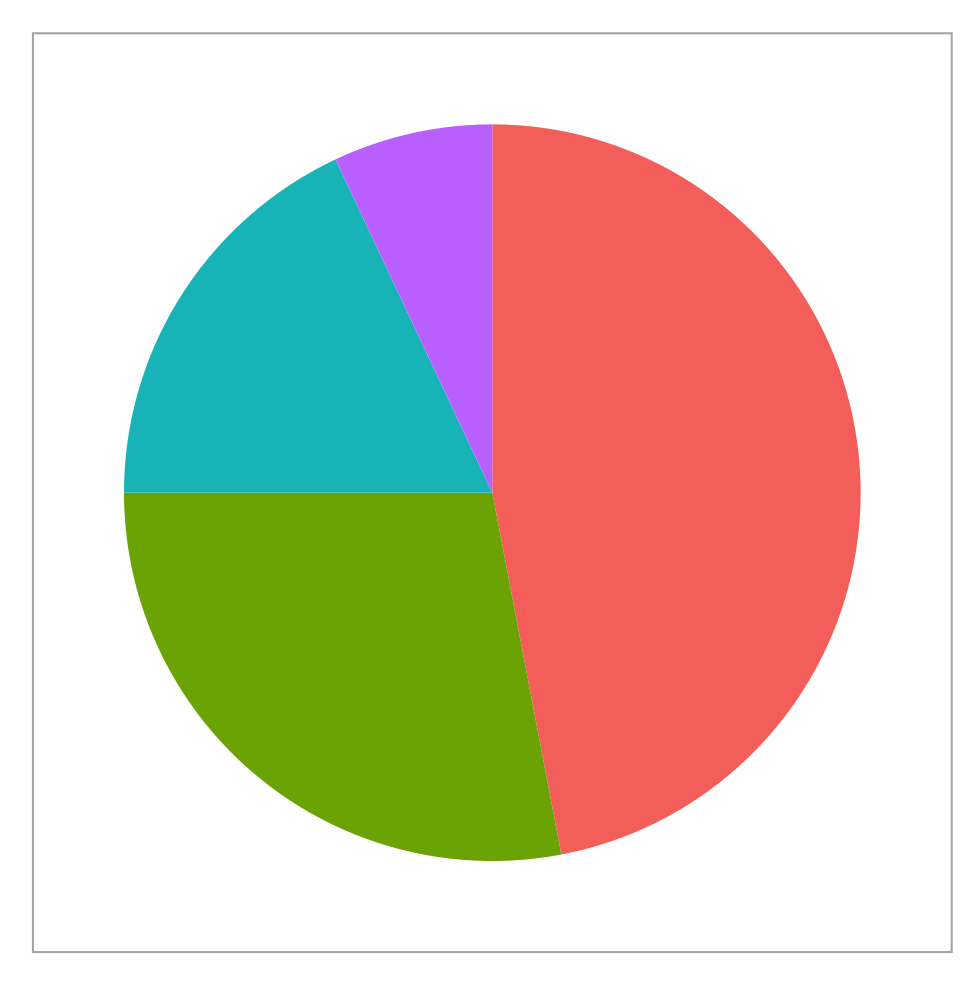} 
        \caption{Empirical Pie Chart}
        \label{fig:comp:cat100:regular}
    \end{subfigure}
    \begin{subfigure}{0.3\linewidth}
        \includegraphics[width=0.9\linewidth]{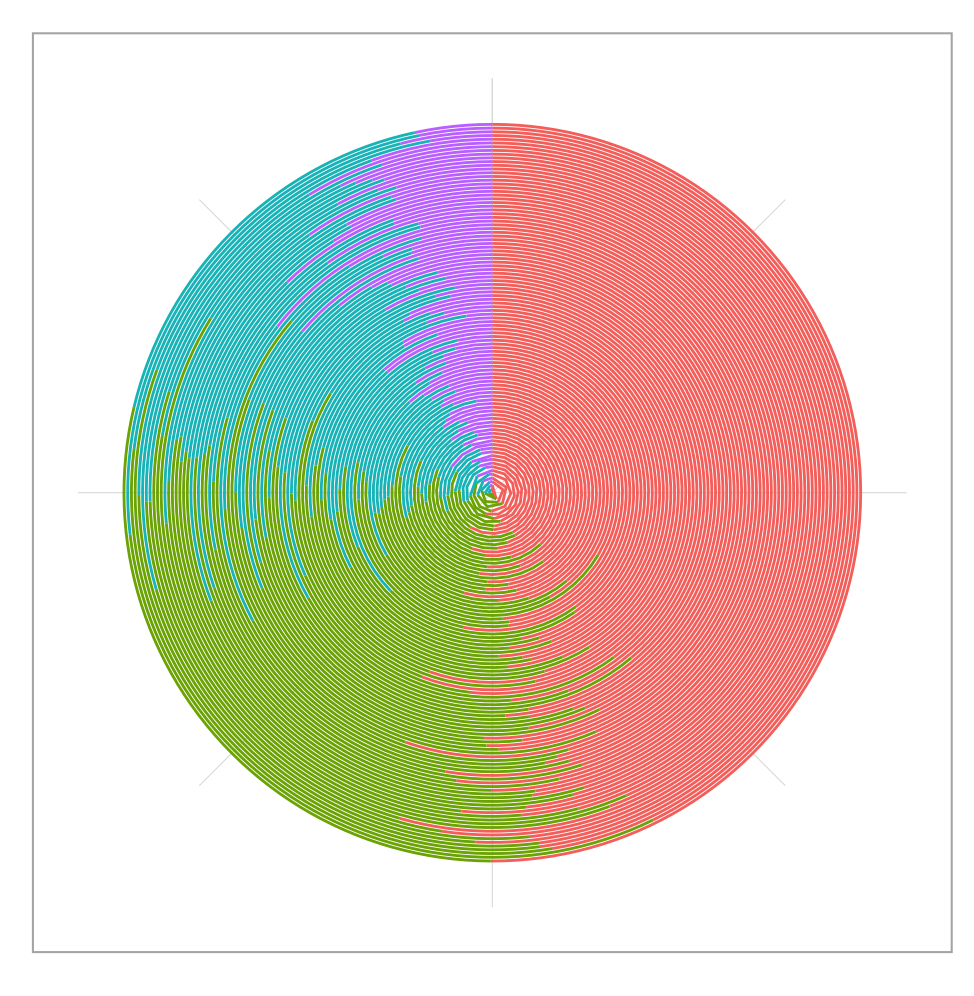} 
        \caption{jk Pie Chart}
        \label{fig:comp:cat100:ours}
    \end{subfigure}

    \begin{subfigure}{0.3\linewidth}
        \includegraphics[width=0.9\linewidth]{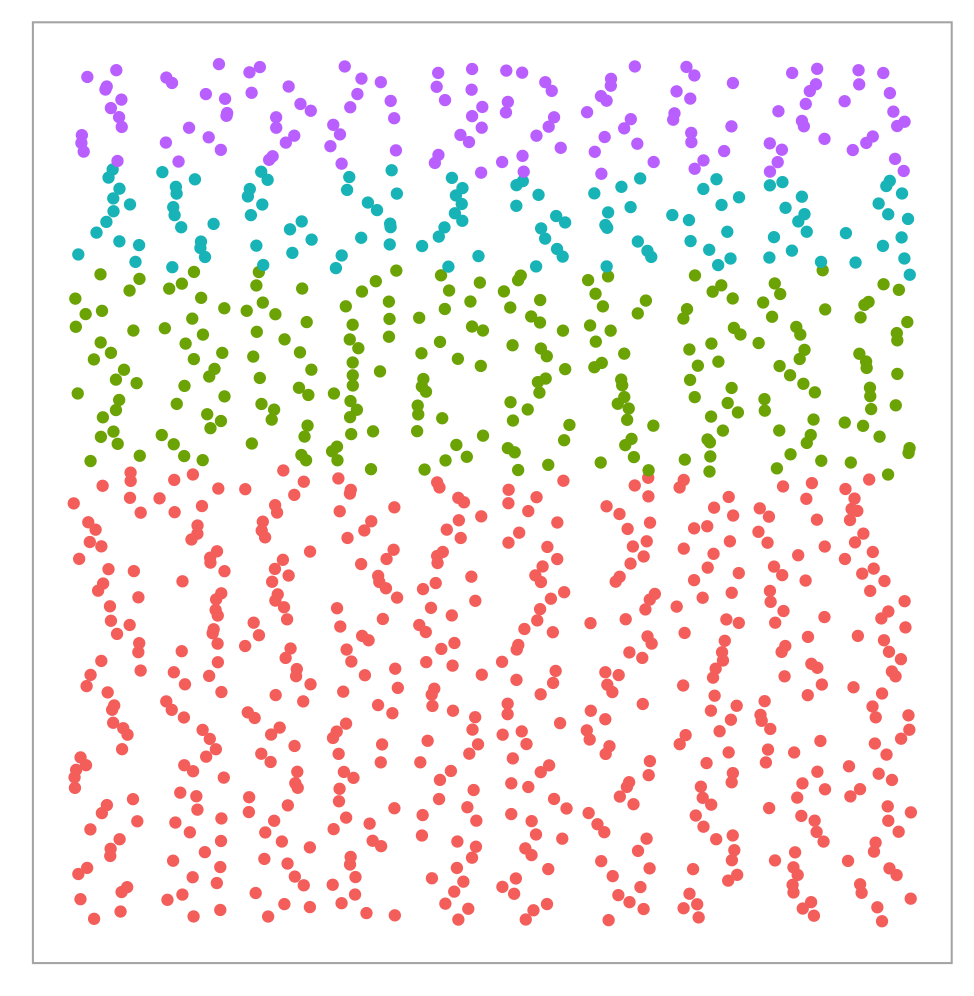} 
        \caption{1000 Samples 
        }
        \label{fig:comp:cat1000:sample}
    \end{subfigure}
    \begin{subfigure}{0.3\linewidth}
        \includegraphics[width=0.9\linewidth]{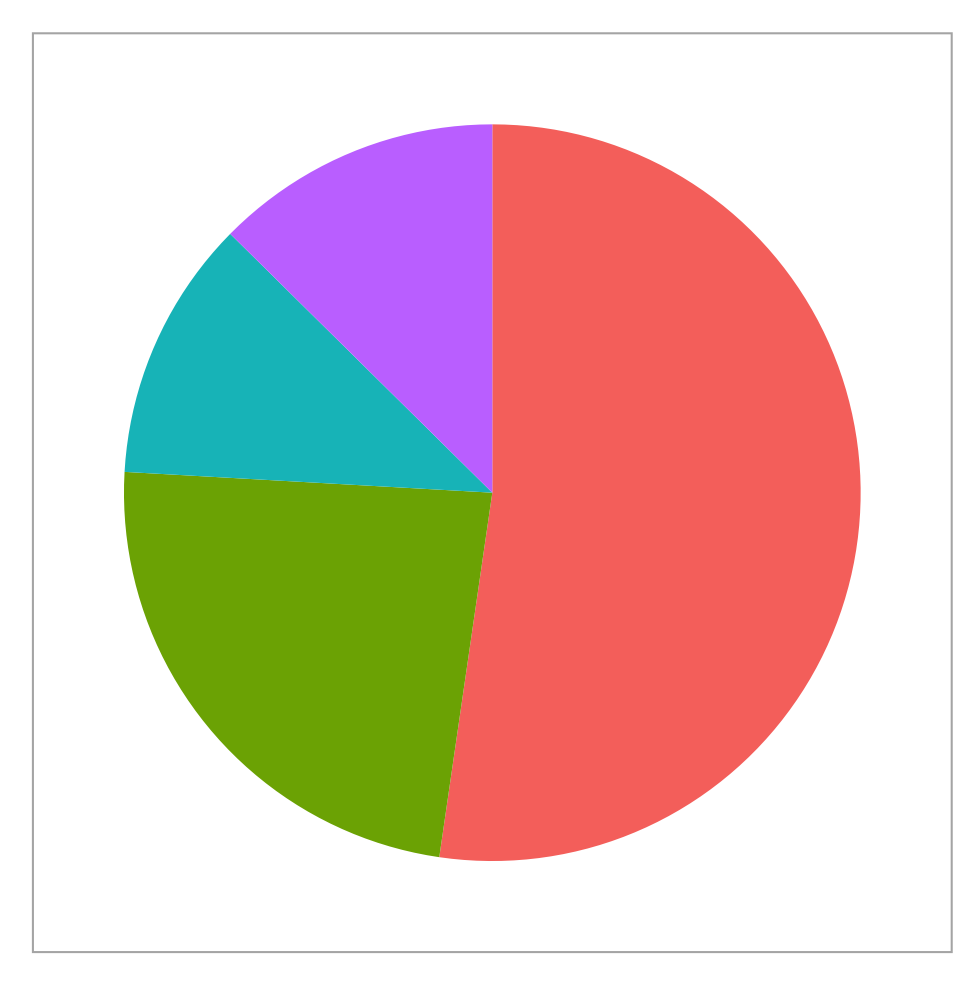} 
        \caption{Empirical Pie Chart}
        \label{fig:comp:cat1000:regular}
    \end{subfigure}
    \begin{subfigure}{0.3\linewidth}
        \includegraphics[width=0.9\linewidth]{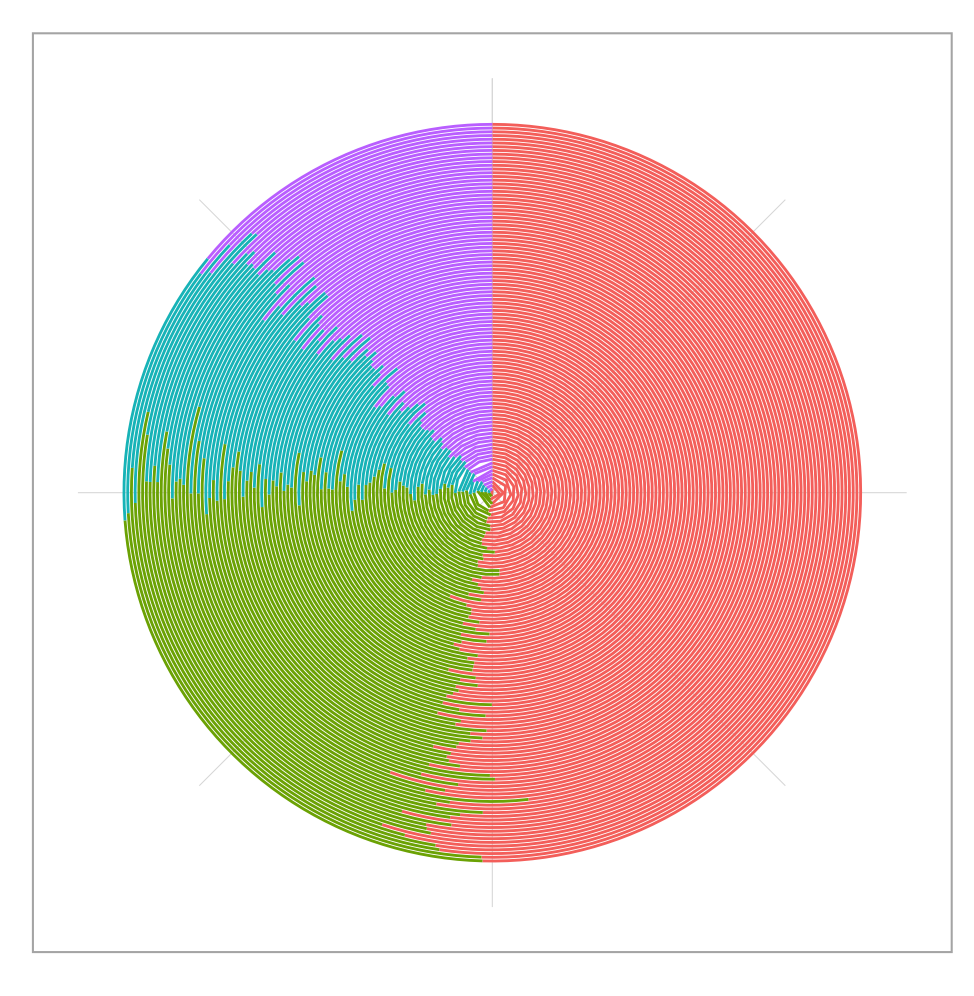} 
        \caption{jk Pie Chart}
        \label{fig:comp:cat1000:ours}
    \end{subfigure}
    
    \caption{Comparison of pie charts for samples drawn from a Multinomial distribution with four classes (proportions 1/2, 1/4, 1/8, 1/8). Rows: 10, 100, and 1000 samples. Columns: Sample, empirical pie chart, jk pie chart with \emph{subset} summary. Subfigure c highlights that class proportions cannot be precisely estimated with the given sample size. In the third row with enough samples, the jk pie chart resembles the empirical pie chart.}
    \label{fig:comp:cat}
\end{figure}

\begin{figure}[tbhp]
    \centering

    \begin{subfigure}{0.3\linewidth}
        \includegraphics[width=0.9\linewidth]{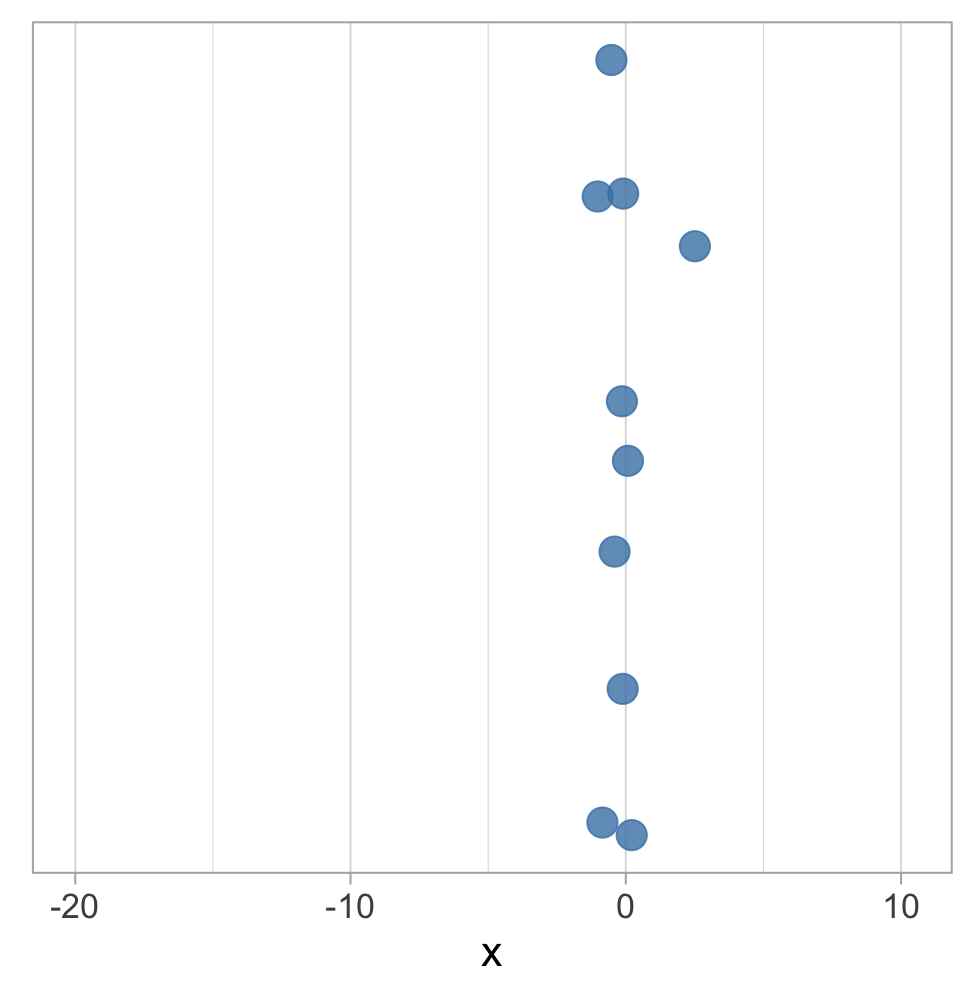} 
        \caption{10 Samples}
        \label{fig:comp:cont10:sample}
    \end{subfigure}
    \begin{subfigure}{0.3\linewidth}
        \includegraphics[width=0.9\linewidth]{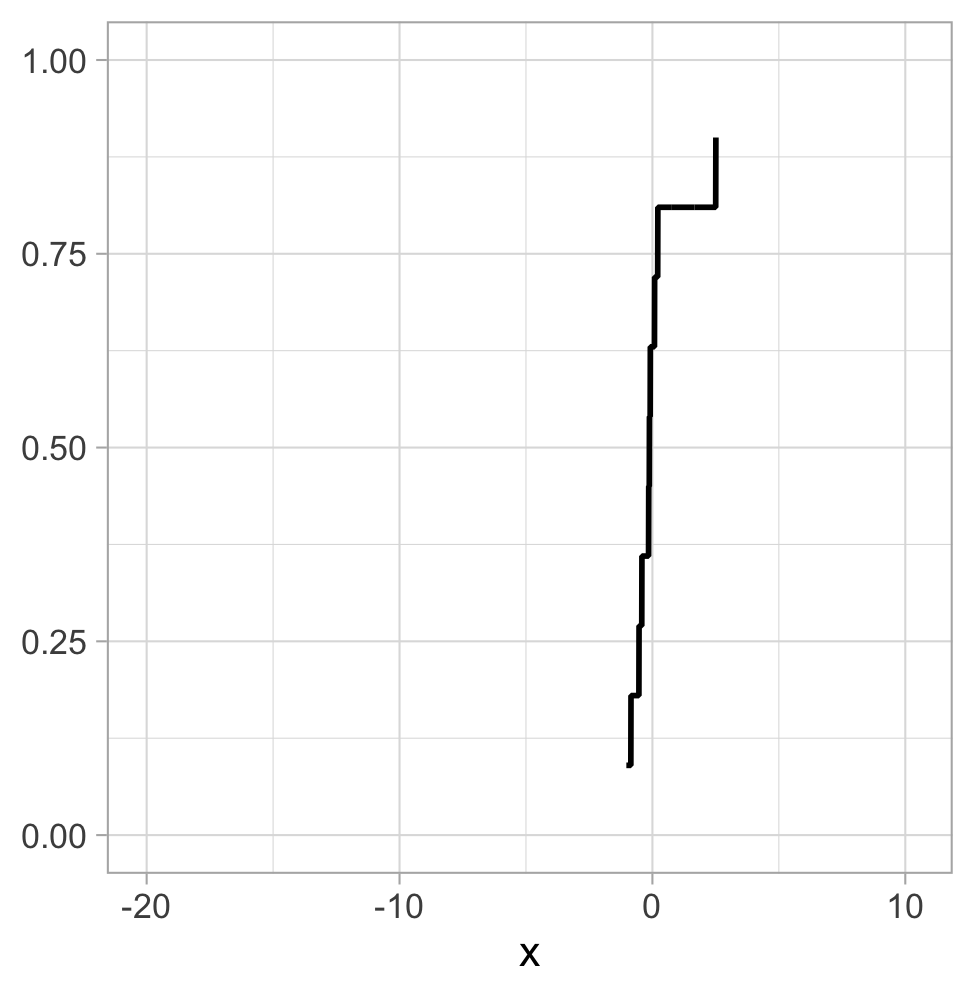} 
        \caption{Empirical CDF}
        \label{fig:comp:cont10:regular}
    \end{subfigure}
    \begin{subfigure}{0.3\linewidth}
        \includegraphics[width=0.9\linewidth]{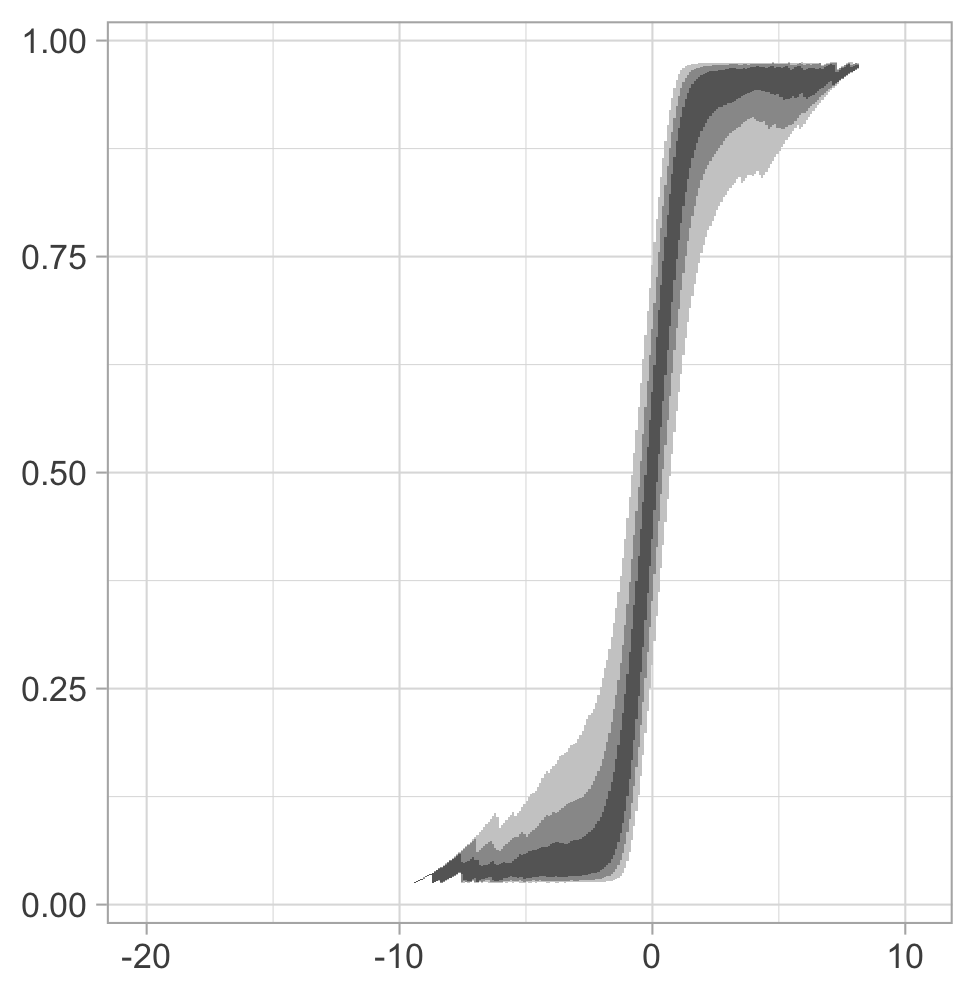} 
        \caption{jk CDF}
        \label{fig:comp:cont10:ours}
    \end{subfigure}

    \begin{subfigure}{0.3\linewidth}
        \includegraphics[width=0.9\linewidth]{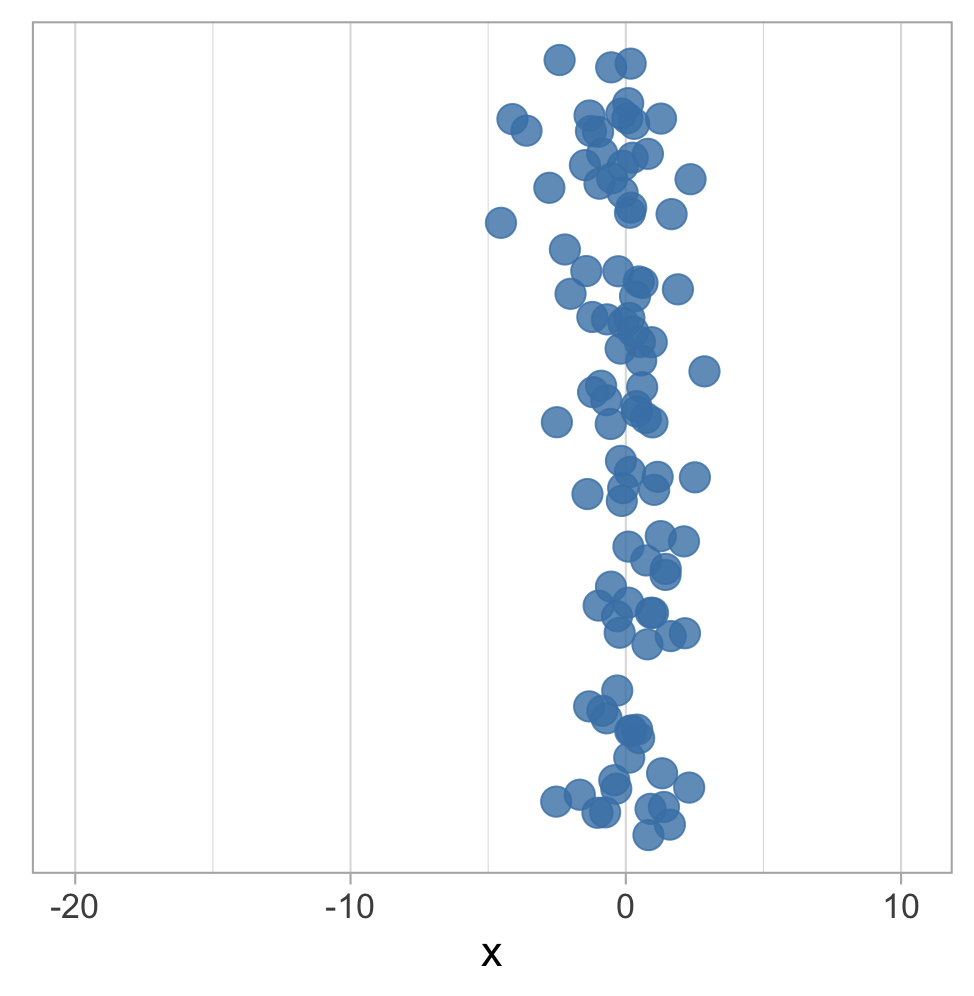} 
        \caption{100 Samples}
        \label{fig:comp:cont100:sample}
    \end{subfigure}
    \begin{subfigure}{0.3\linewidth}
        \includegraphics[width=0.9\linewidth]{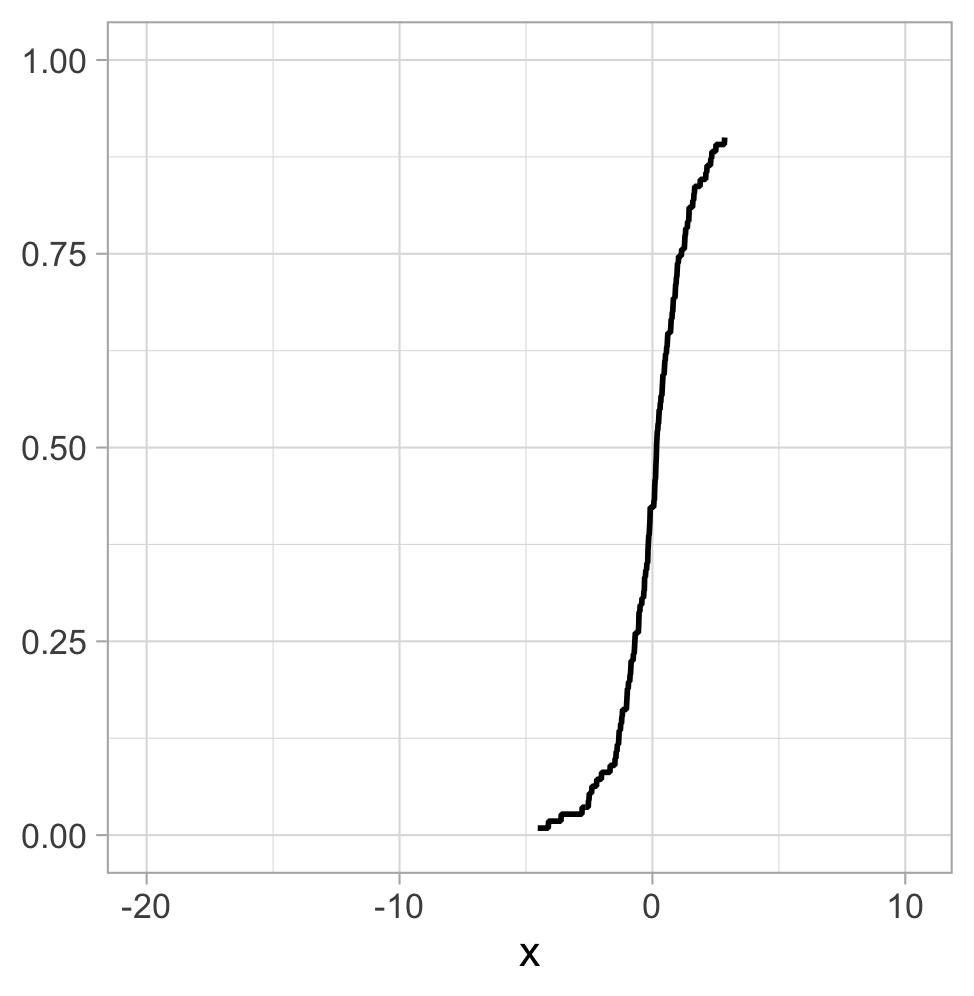} 
        \caption{Empirical CDF}
        \label{fig:comp:cont100:regular}
    \end{subfigure}
    \begin{subfigure}{0.3\linewidth}
        \includegraphics[width=0.9\linewidth]{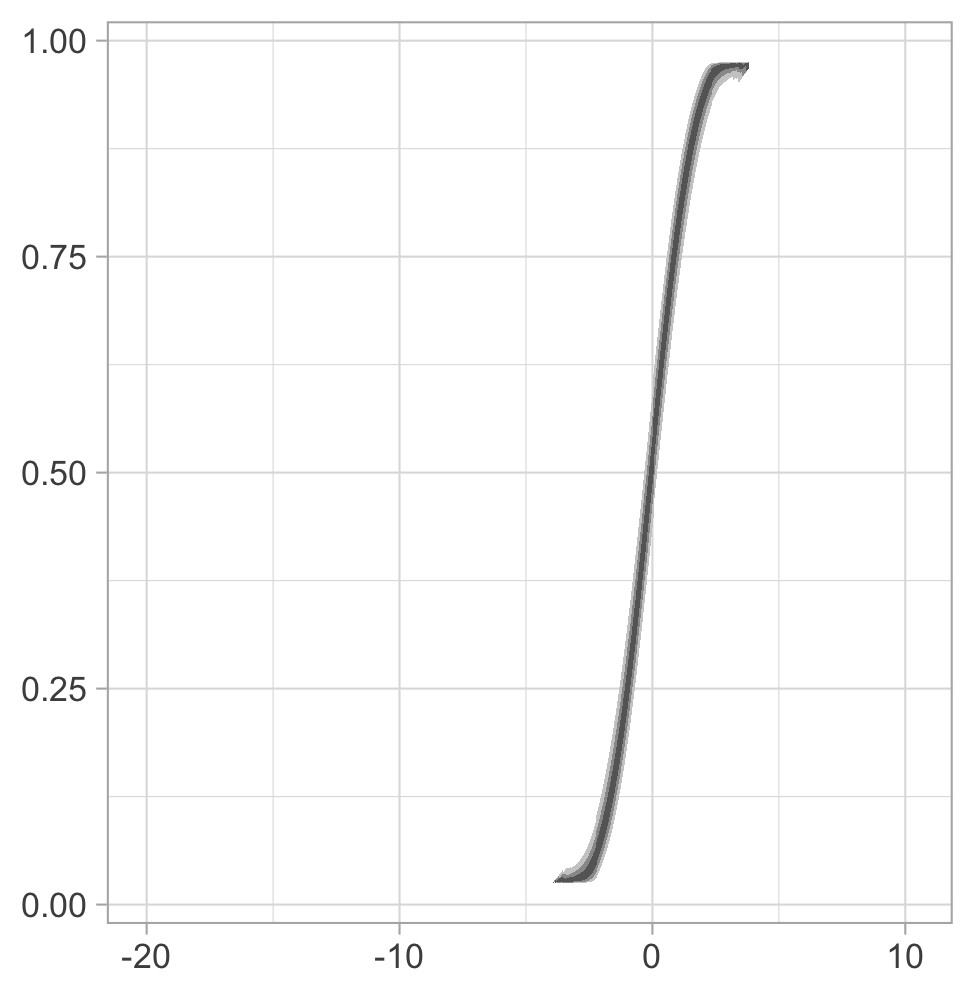} 
        \caption{jk CDF}
        \label{fig:comp:cont100:ours}
    \end{subfigure}

    \begin{subfigure}{0.3\linewidth}
        \includegraphics[width=0.9\linewidth]{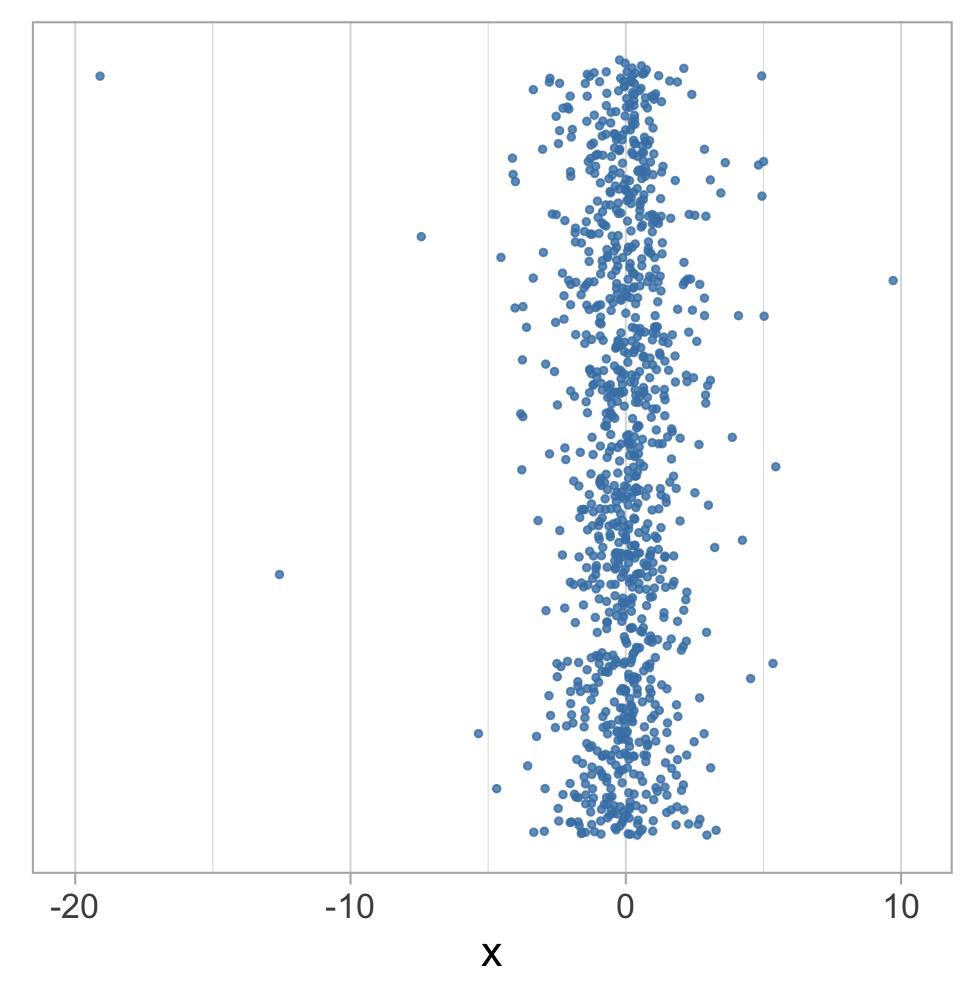} 
        \caption{1000 Samples}
        \label{fig:comp:cont1000:sample}
    \end{subfigure}
    \begin{subfigure}{0.3\linewidth}
        \includegraphics[width=0.9\linewidth]{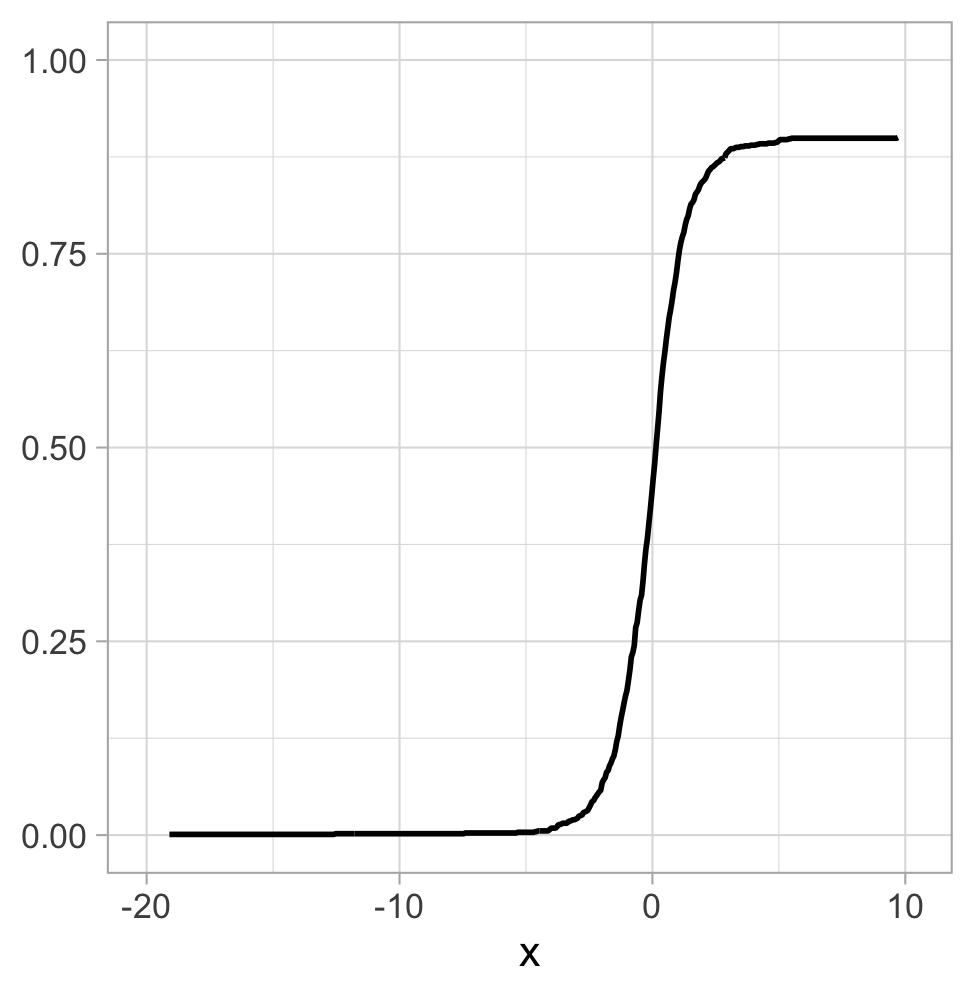} 
        \caption{Empirical CDF}
        \label{fig:comp:cont1000:regular}
    \end{subfigure}
    \begin{subfigure}{0.3\linewidth}
        \includegraphics[width=0.9\linewidth]{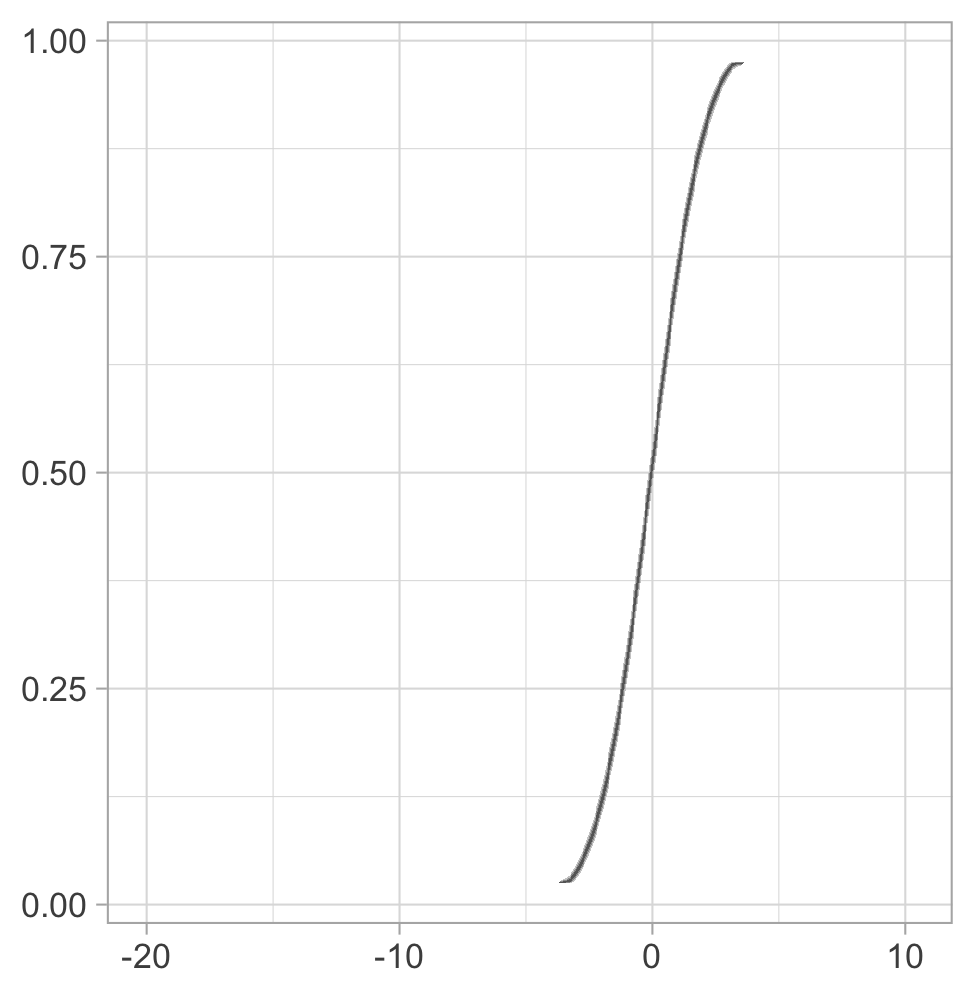} 
        \caption{jk CDF}
        \label{fig:comp:cont1000:ours}
    \end{subfigure}

    \caption{
        Comparison of CDF plots for samples drawn from a Student(3, 0, 1) distribution. Rows: 10, 100, and 1000 samples. Columns: Sample, empirical CDF (eCDF), jk CDF based on a Normal model with \emph{band} summary. The empirical plot was generated with \texttt{ggdist}. The eCDF always shows crisp numbers, while the jk CDF gets more detailed with more samples.
    }
    
    \label{fig:comp:cont}
\end{figure}

\begin{figure}[tbhp]
    \centering

    \begin{subfigure}{0.3\linewidth}
        \includegraphics[width=0.9\linewidth]{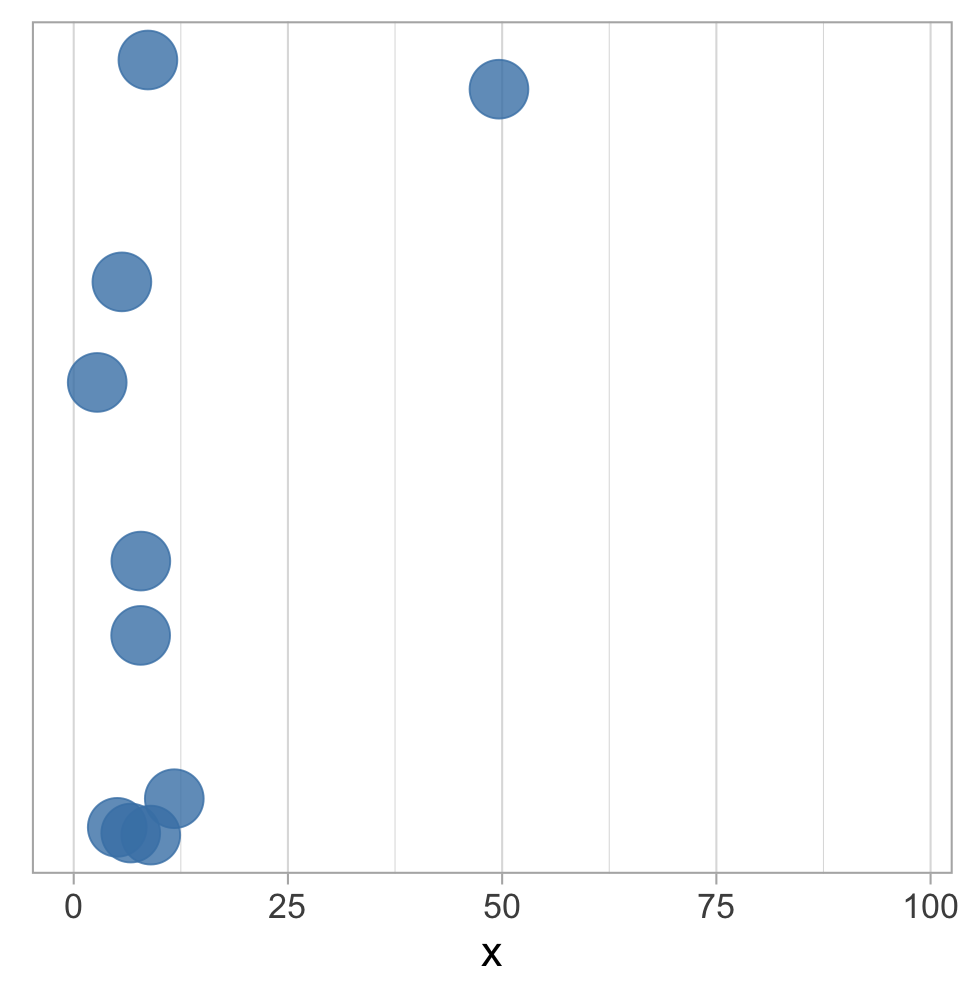} 
        \caption{10 Samples}
        \label{fig:comp:skew10:sample}
    \end{subfigure}
    \begin{subfigure}{0.3\linewidth}
        \includegraphics[width=0.9\linewidth]{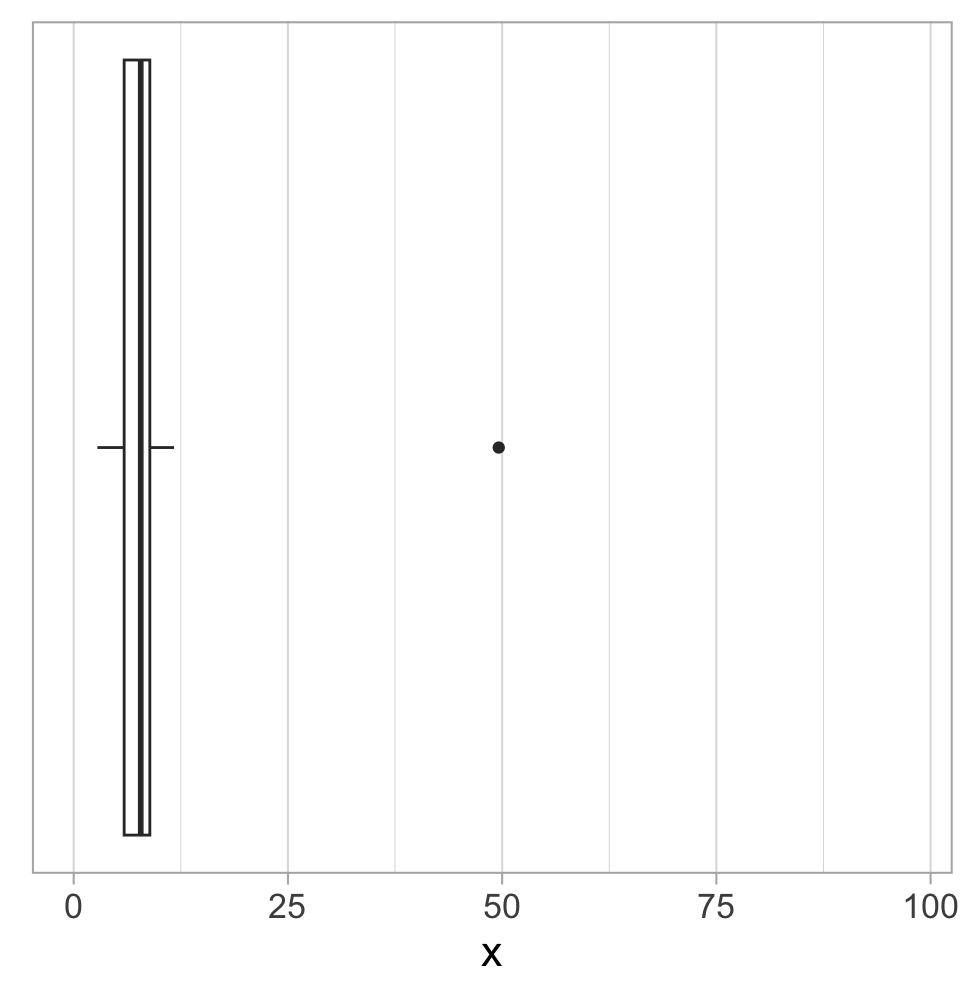} 
        \caption{Empirical Boxplot}
        \label{fig:comp:skew10:regular}
    \end{subfigure}
    \begin{subfigure}{0.3\linewidth}
        \includegraphics[width=0.9\linewidth]{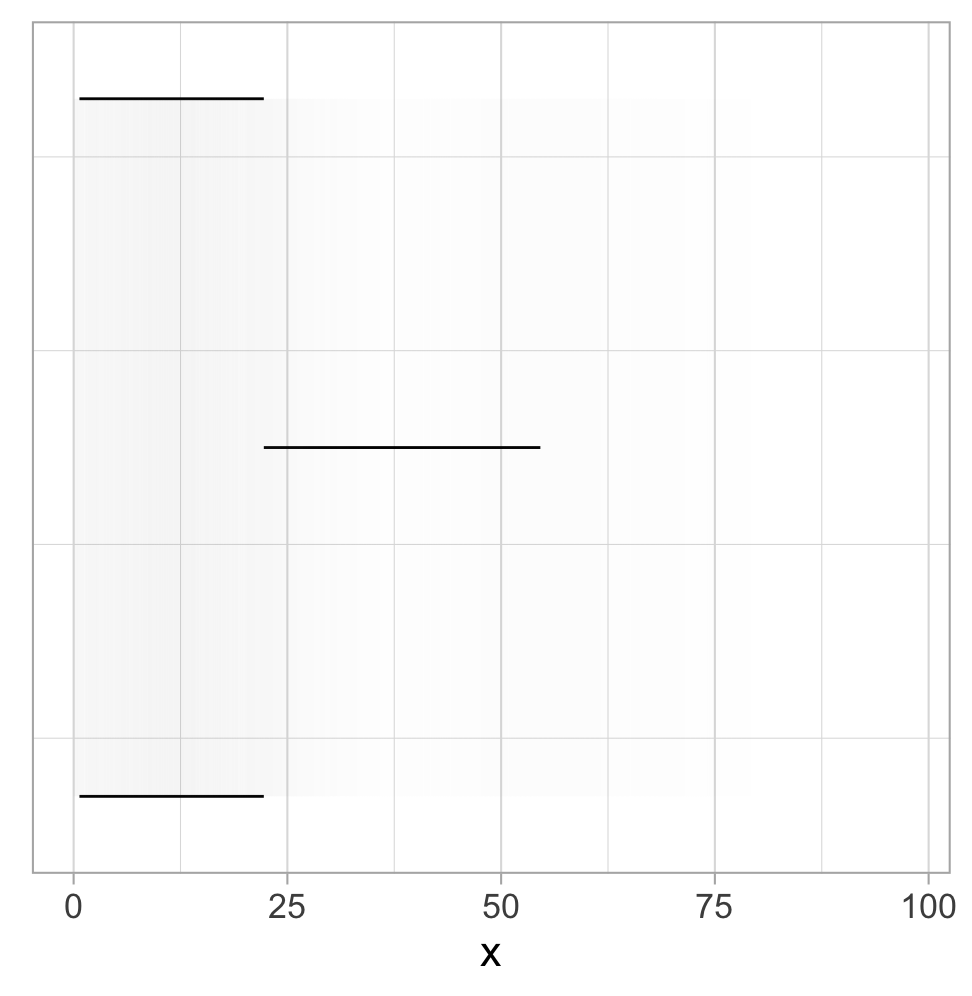} 
        \caption{jk Boxplot}
        \label{fig:comp:skew10:ours}
    \end{subfigure}

    \begin{subfigure}{0.3\linewidth}
        \includegraphics[width=0.9\linewidth]{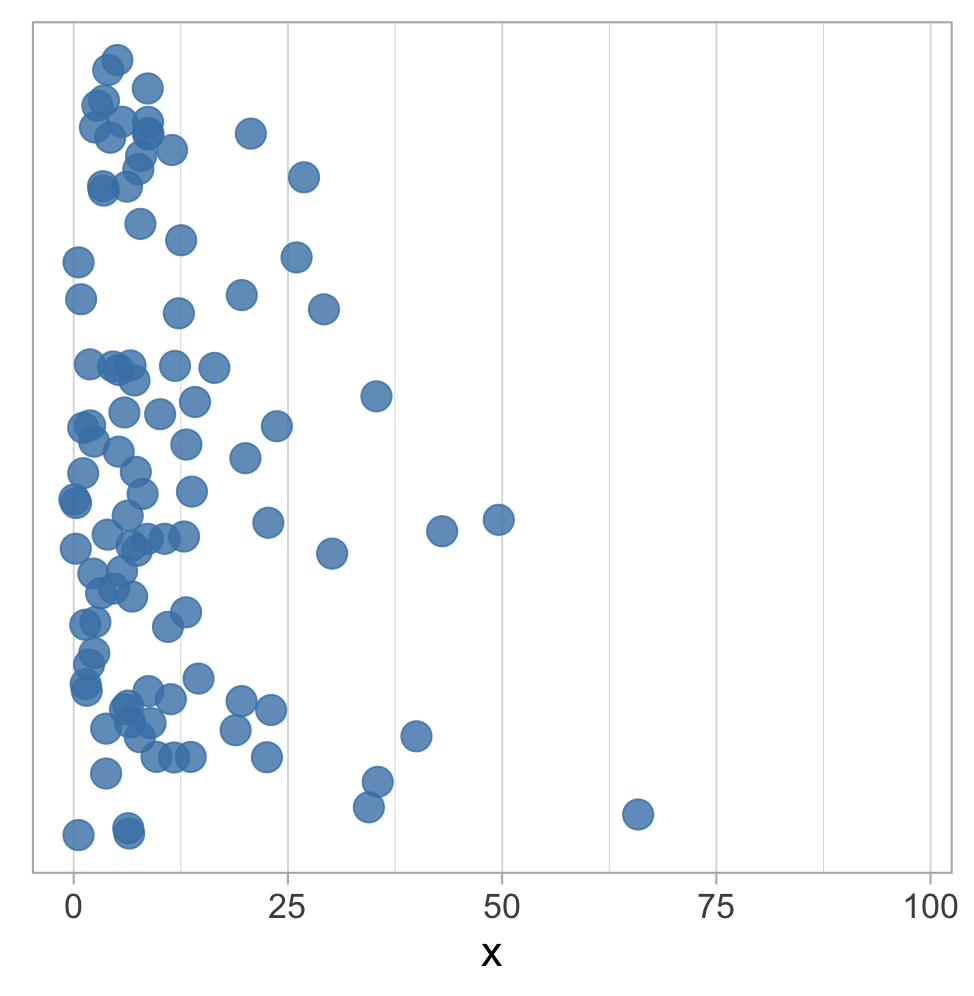} 
        \caption{100 Samples}
        \label{fig:comp:skew100:sample}
    \end{subfigure}
    \begin{subfigure}{0.3\linewidth}
        \includegraphics[width=0.9\linewidth]{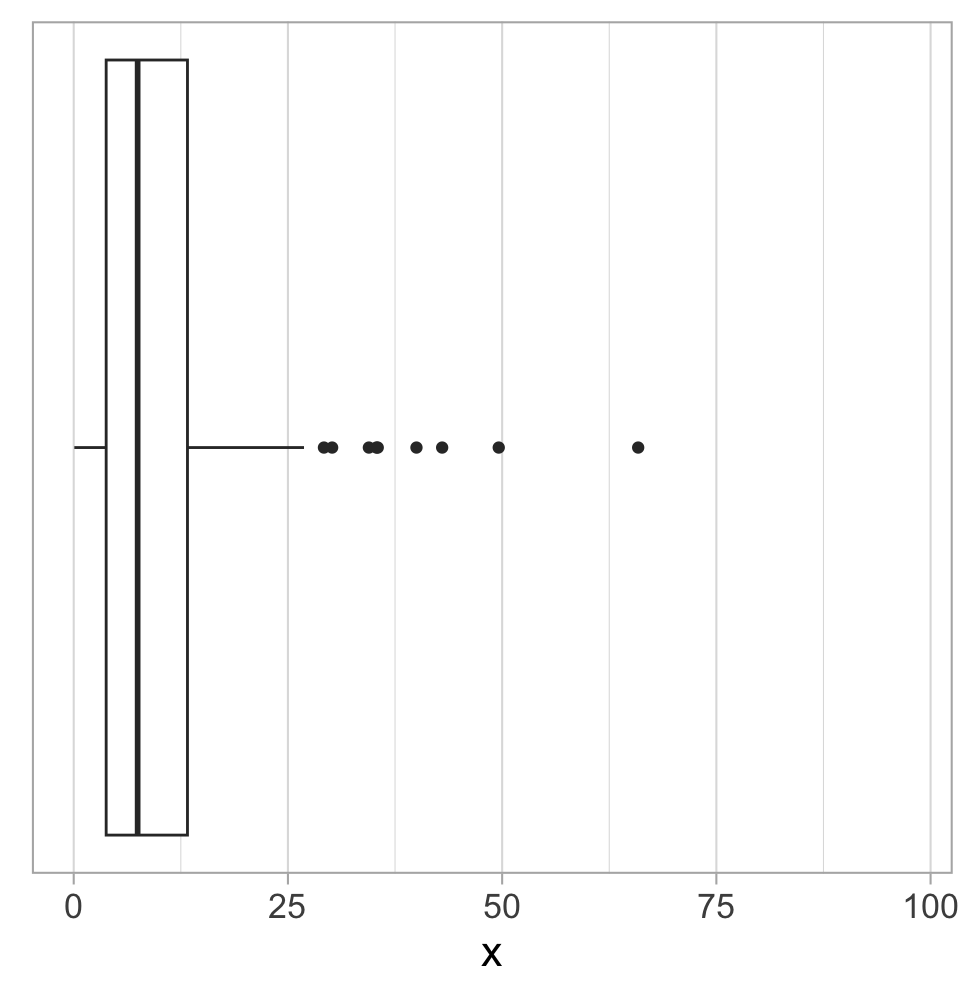} 
        \caption{Empirical Boxplot}
        \label{fig:comp:skew100:regular}
    \end{subfigure}
    \begin{subfigure}{0.3\linewidth}
        \includegraphics[width=0.9\linewidth]{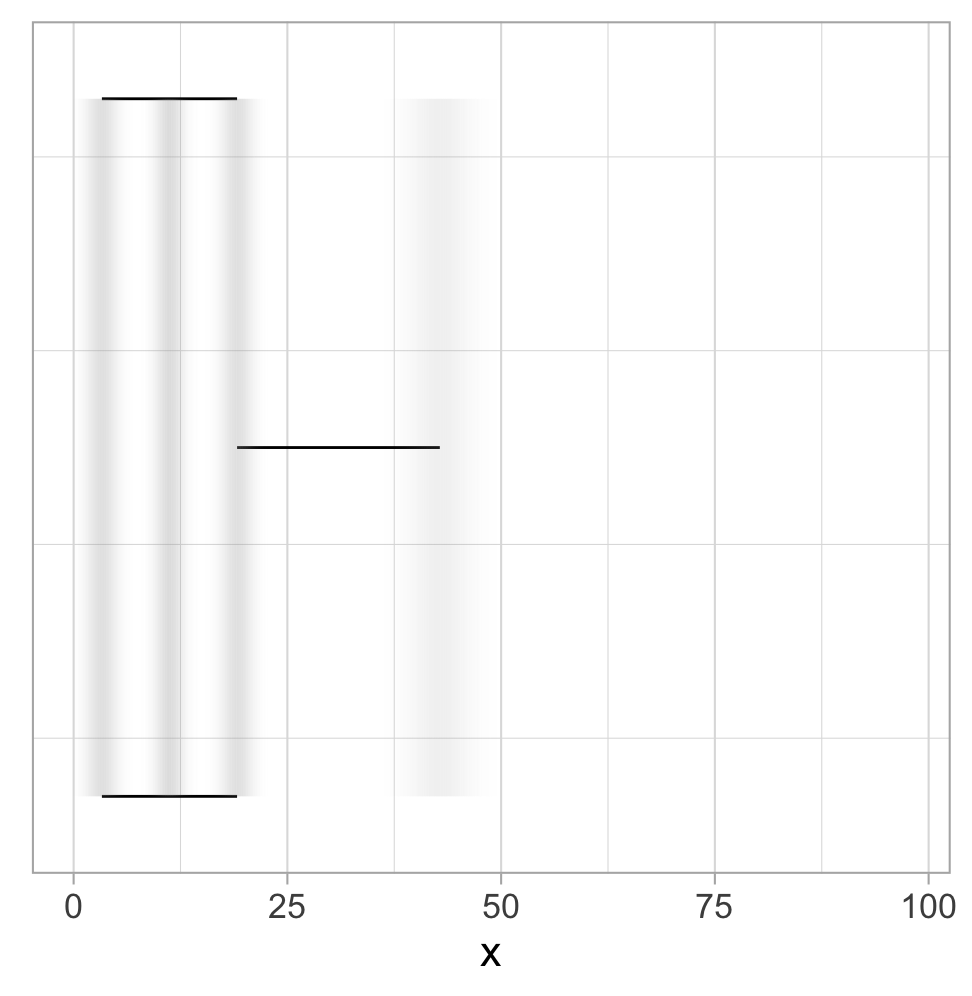} 
        \caption{jk Boxplot}
        \label{fig:comp:skew100:ours}
    \end{subfigure}

    \begin{subfigure}{0.3\linewidth}
        \includegraphics[width=0.9\linewidth]{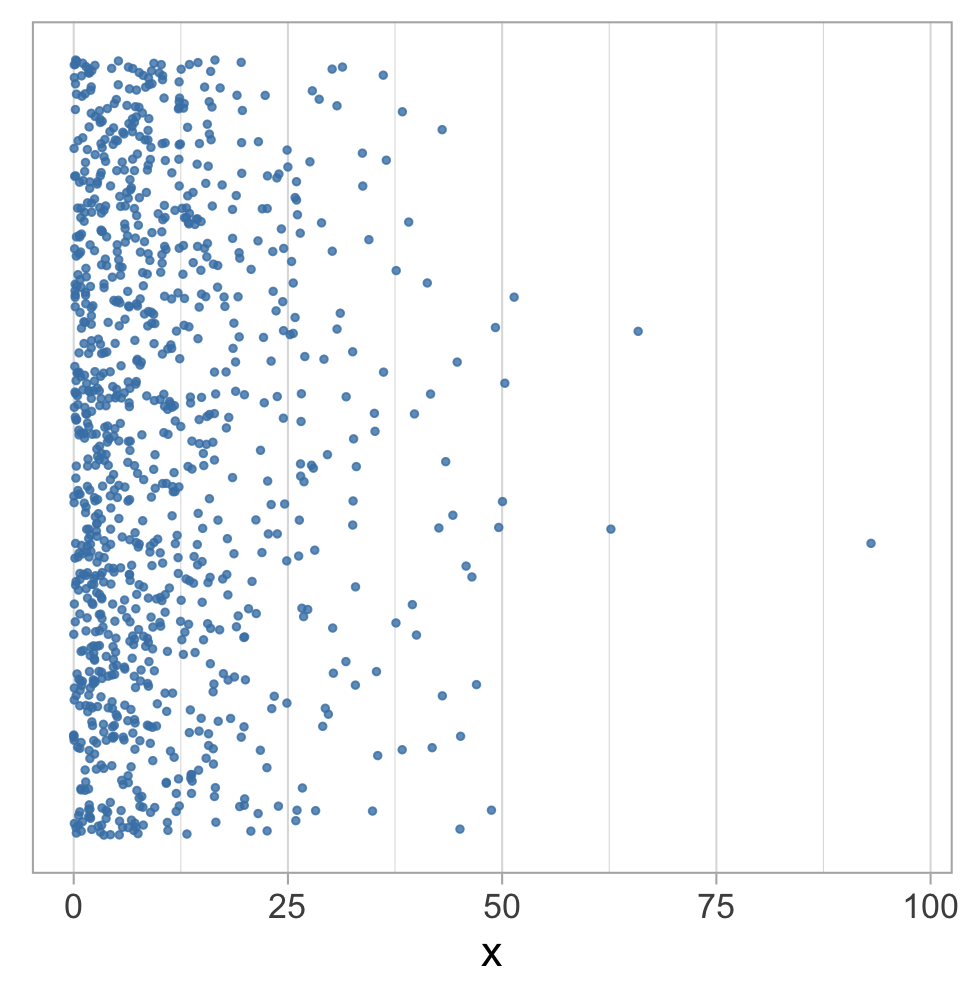} 
        \caption{1000 Samples}
        \label{fig:comp:skew1000:sample}
    \end{subfigure}
    \begin{subfigure}{0.3\linewidth}
        \includegraphics[width=0.9\linewidth]{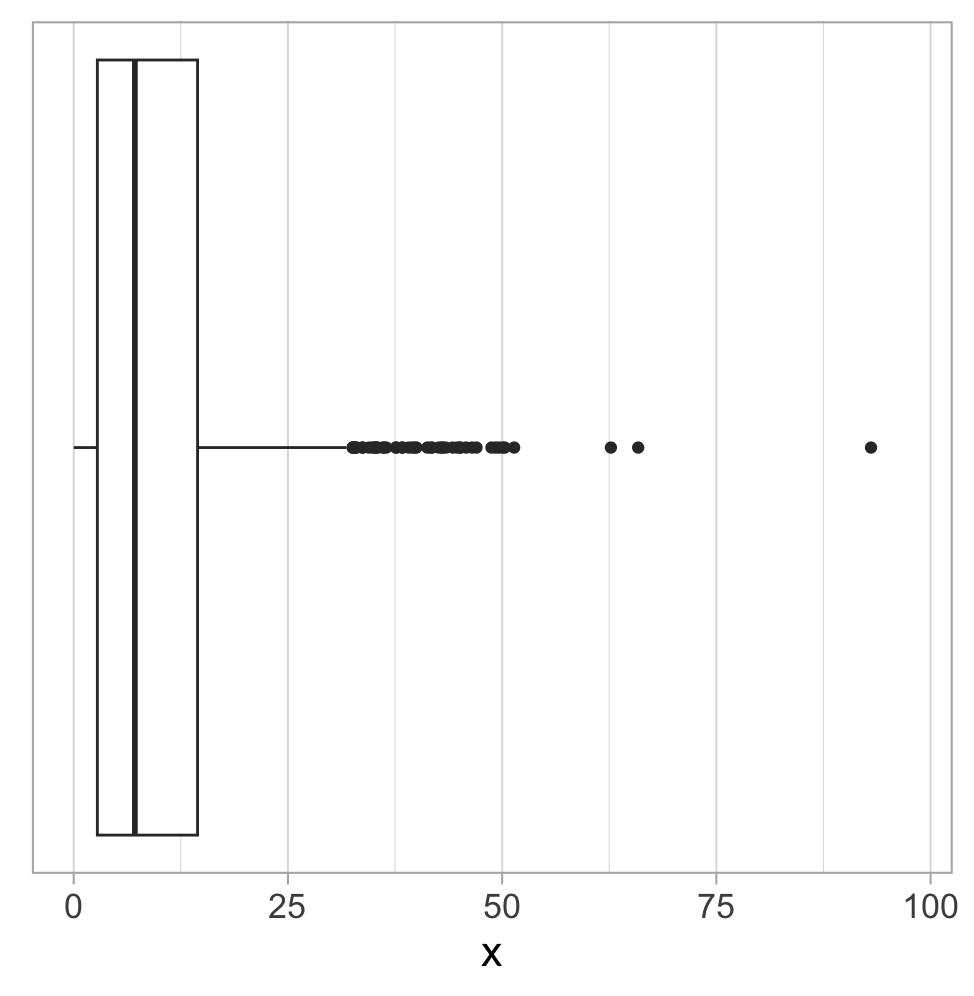} 
        \caption{Empirical Boxplot}
        \label{fig:comp:skew1000:regular}
    \end{subfigure}
    \begin{subfigure}{0.3\linewidth}
        \includegraphics[width=0.9\linewidth]{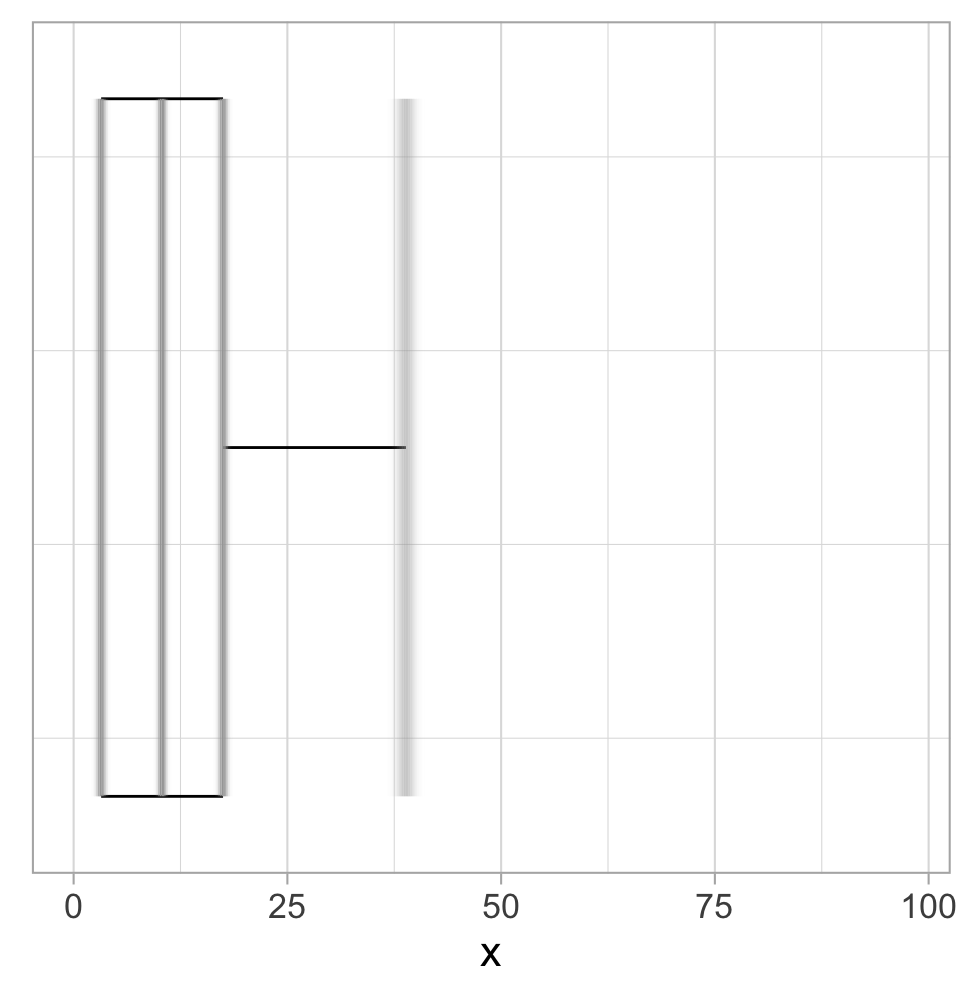} 
        \caption{jk Boxplot}
        \label{fig:comp:skew1000:ours}
    \end{subfigure}
    
    \caption{
        Example of a mismatch between data and model. Samples are drawn from a skewed and left-bounded Gamma(1, 1/10) distribution. Rows: 10, 100, and 1000 samples. Columns: Sample, empirical boxplot, jk boxplot based on a Normal model with \emph{gradient} summary. The empirical boxplot (generated with \texttt{ggplot2}) more closely resembles the sample distribution, as it is based on descriptive statistics.}
    
    \label{fig:comp:skew}
\end{figure}

\paragraph{Data Generation.}  We consider three sample sizes with different orders of magnitude ($n \in \{10, 100, 1000\}$), all using the same random seed.
For categorical data, we drew one assignment of $n$ items into four classes using \texttt{rmultinom(1, n, p)} (\cref{fig:comp:cat}), with proportions ($p$) 1/2, 1/4, 1/8, and 1/8, respectively.
For continuous data, we drew $n$ observations from a Student distribution with parameters $\nu=3, \mu=0, \sigma=1$ (\cref{fig:comp:cont}), and a Gamma distribution with parameters $\alpha=1, \theta=1/10$ (\cref{fig:comp:skew}).

\paragraph{Categorical Data.} \cref{fig:comp:cat} shows the three samples (a, d, g), empirical pie charts of the samples with standard \texttt{ggplot2} techniques (b, e, h), and the corresponding jk-plots (c, f, i) with \emph{subset} summary (100 draws). The population proportions $p$ are the same in all subfigures. 
For $n=10$ (a--c), the empirical pie chart looks perfectly certain, and shows the green class with crisp proportion $1/10$. The jk pie chart, on the other hand, excludes only very small proportions of the red class while being very uncertain about others. The sample proportions could potentially be added to the jk pie chart using an overlay. 
Increasing the sample size further ($n=100$, d--f) does not eliminate the problem of the empirical pie chart misrepresenting population proportions. The jk pie chart now resembles its empirical counterpart more closely, but includes the true parameters as some of the concentric donuts cover the true proportions.
The same holds for the biggest sample ($n=1000$, g--i), where the jk pie chart is very close to the empirical one, but still some reference models (e.g., green rings going across 6 o'clock) include the true proportions.

In summary, the jk pie chart appropriately communicates the uncertainty of population proportion estimates relative to the small sample size (P1) and fluidly adapts its visual representation (P2). As a consequence, the jk pie chart limits extrapolations of trends and patterns from small to larger samples.
The crisp boundary between red and purple is a rendering artifact of our chart implementation as horizontal stacked bars in a polar coordinate system. It could be remedied by optimizing the alignment of arcs' centers~\cite{sun2023}.


\paragraph{Continuous Data.} \cref{fig:comp:cont} shows the three samples (a, d, g), empirical CDFs (eCDF; b, c, h), and jk CDFs with \textit{band} summary (c, f, i). Here, the sample-generating distribution (Student) somewhat matches the jk CDF's assumption of a Normal distribution.
Contrary to the categorical example, the empirical CDF for $n=10$ (a--f) looks somewhat broken (no tails, staircase pattern), thus already hinting at a too-small sample (P2). Nevertheless, a mean (at $y=.5$) may be read off the plot perfectly well (P1). The jk CDF shows plausible CDFs and again limits unwarranted extrapolation as plausible means are aggregated into a band (95\% interval $-.75,.75$).
With more data ($n \in \{100, 1000\}$, d--i), the jk CDF approaches the eCDF. As in the categorical case, the jk CDF quantifies the uncertainty introduced by sample size accurately (P1) and visually adapts nicely to the sample (P2).

One assumption of jk-plots is that their model matches the data. 
We demonstrate what happens when this assumption is violated in \cref{fig:comp:skew}. The samples come from a Gamma distribution, which is skewed and bounded, while the default Normal model can only handle symmetric and unbounded data. The center column in \cref{fig:comp:skew} (b, e, h) shows empirical boxplots, while the right column (c, f, i) shows the jk counterparts with \textit{gradient} summary. While the jk boxplot adapts the visual accuracy of its marks to the sample size and the empirical boxplot does not, the latter more accurately reflects the skewness in the distribution (visible by the offset center tick and outlier points outside the whiskers). 
We implemented jk boxplots with gradients such that they represent the distribution of summary statistics defined by the posterior. 
This meant finding a compromise between a visual idiom resembling a boxplot and accurate visual encoding. 
We opted to hide the outliers, but another feasible strategy would be to show them if they are outside the whiskers' 95\% interval. Horizontal lines extend to the summaries' center estimate, but they could also be gradually subdued. 

\section{Interviews}
\label{sec:interviews}


To understand how our proposed uncertainty visualizations fit into real-world data analysis workflows and visualization applications, we conducted semi-structured interviews with experts from various fields, discussing their practices and needs related to small sample sizes and the potential benefits and challenges of jk-plots. 

The interviews were split into two parts. First, we asked participants about visualization tasks and characteristics of small samples relevant to their work.
In the second half, we showed them previously prepared jk-plots, explained the technical workflow, and discussed the results.
Our interviewees comprised researchers in statistics (\textbf{S1}) and visualization (\textbf{V1}, \textbf{V2}), and an applied data scientist (\textbf{D1}).
As this work was partly motivated by the grievances of a co-author, who regularly works with small samples, they also provided feedback from their work (\textbf{V3}).
Interviews lasted 30--60 minutes and were conducted via Zoom. One co-author led the interview while one or two co-authors took notes.

\paragraph{Small Sample Use Cases.}
All participants had encountered datasets with few observations.
\textbf{V1} and \textbf{V2} work on dynamic networks in art history, especially art exhibitions. Few observations may appear in the form of small collectives, comprised of few artists (3--10). Both the relevant data for each artist and the network may change over time. 
Similarly, \textbf{V3} often conducts lab studies with relatively few participants (ca. $30$), collecting several quantitative data points per participant.
\textbf{S1} works mostly on multivariate and functional data, so for them, a single observation could still mean several data points. They mentioned a possible needle-in-haystack situation where few relevant observations are hidden in lots of noise. \textbf{D1} reported working with synthetic data and with performance indicators of technical compounds, measured at few time points. 

The experts' goals ranged from model-free data exploration (\textbf{V1}, \textbf{V2}), to communication of sample statistics to scientific peers (\textbf{V3}), to statistical inference from the available sample (\textbf{S1}, \textbf{D1}), as well as communicating those results to stakeholders (\textbf{D1}). Common tasks included the comparison of two or more samples of different groups (\textbf{V1--V3}), outlier detection (\textbf{S1}), and assessing the fit of statistical models (\textbf{D1}).
Regarding current visualization strategies for small samples, experts reported using univariate visualizations ranging from plotting raw data points (\textbf{D1}) to summary visualizations such as boxplots (\textbf{V3}) to visualizations of model parameters based on the data (\textbf{D1}). These choices are related to aesthetic preferences (``plotting raw data feels unfinished'', \textbf{V3}), the convenience of charts easily available in commonly used tools (\textbf{V3}), and an aim for transparently displaying the data (\textbf{D1}).

\paragraph{Feedback on jk-plots.} All experts found jk-plots an interesting approach, but raised some concerns regarding their usability, accuracy, and adaptation. Statistics experts (\textbf{S1}, \textbf{D1}) highlighted the need for a proper model and prior distributions, along with sensitivity analysis of the latter, an issue discussed at length in the literature~\cite{gelman2020, kruschke2015, kruschke2021, mcelreath2022}. Coming from a frequentist statistics background, both were uncomfortable with the involved subjectivity and a potential lack of reproducibility.
Used to formulating statistical models to address their problems, \textbf{S1} and \textbf{D1} were unclear on the role of jk-plots in data analysis, which can be done more flexibly without them, and instead saw their main purpose in result communication and obtaining a ``general understanding of the data.''

All experts were hesitant about jk-plots for audiences with lower statistic literacy, such as clients in an applied data science consulting setting (\textbf{D1}), researchers in the humanities (\textbf{V1}, \textbf{V2}), or the general public (all). They questioned both the understandability for these audiences, and the audiences' need of understanding distribution parameters. To the best of our knowledge, the empirical research into that is missing. In particular, \textbf{V1} and \textbf{V2} expected their audiences to understand the visual idioms, but not to be familiar with or interested in the specific statistical values displayed. 

However, \textbf{D1} saw potential as an uncertainty communication tool, highlighting that even though lay audiences may not understand all aspects of the jk-plot in detail, the plot effectively conveys the uncertainty. 
\textbf{V1} and \textbf{V2} mentioned that an adapted version of the jk-plot might help contextualize the reliability of an art style detection model. 
\textbf{V3} appreciated how the blurry boxplot in particular acknowledges the uncertainty for the encoded summary statistics by showing a range rather than a potentially highly inaccurate estimation. Considering they already use boxplots to display statistical characteristics of their small samples, this points to the jk-plot as a valuable alternative. 

Improvement suggestions for jk-plots included highlighting the most likely estimate, while acknowledging that this could be detrimental to the audience if they choose to ignore the sample size uncertainty (\textbf{D1}), or overlaying descriptive statistics of the sample (\textbf{S1}, \textbf{D1}). \textbf{V1} and \textbf{V2} were interested in a potential adaptation of the jk-plots to other data types, such as temporal, spatial, or network data. They further suggested adapting jk-plots to use visual idioms to specifically address less statistically inclined audiences. 
As a possible solution to the model--data discrepancy, in particular the case of multimodality or skewness (cf. \cref{fig:comp:skew}), \textbf{S1} suggested implementing a Gaussian mixture model.


\section{Discussion}

In this paper, we investigated jk-plots as an uncertainty visualization technique that conveys the uncertainty arising from the number of available observations. They emulate familiar encoding idioms (e.g., box plots or pie charts) and their visual encoding becomes more crisp with more evidence. 
We compared jk-plots to their commonly known counterparts (\cref{sec:examples}) and gathered the expert opinion of practitioners (\cref{sec:interviews}). 

In the process, we identified both opportunities and challenges. An interesting open question, for instance, is whether jk-plots can be accurately interpreted by a lay audience and whether the plots modulate the viewer's confidence. We plan to conduct such an experiment in the future. It is also not clear whether lay people perceive \enquote{broken} uncertainty visualizations (e.g., \cref{fig:comp:cont10:regular}) as less trustworthy than \enquote{whole} ones (e.g., \cref{fig:comp:cont1000:regular}). Experts commented on jk-plots' lack of versatility, which may be countered by sophisticated uncertainty elicitation techniques~\cite{hullman2018, kim2021, kim2019, piccolotto2025a} to incorporate prior knowledge/beliefs. Regarding the model formulation, one could allow to select from more distributions (e.g., Skew Normal to account for skewness, Student for outliers), but at some point, analysts will profit more from a custom model. Finally, there are limits to the jk-plots' adaptivity to the sample size, as the model may become under-specified (more parameters than observations) and techniques to set data-driven priors may not be applicable for very few observations.

\acknowledgments{%
	This research was funded in whole or in part by the Austrian Science Fund (FWF) [10.55776/PAT6437224]. For open access purposes, the author has applied a CC BY public copyright license to any author-accepted manuscript version arising from this submission.
}

\bibliographystyle{abbrv-doi}

\bibliography{template}

\end{document}